\documentclass{lmcs}
\pdfoutput=1

\usepackage{lastpage}
\lmcsdoi{17}{3}{16}
\lmcsheading{}{\pageref{LastPage}}{}{}%
{Nov.~04,~2020}{Aug.~10,~2021}{}


\usepackage[british]{babel}
\usepackage[utf8]{inputenc}
\usepackage[T1]{fontenc}

\newcommand{\R}{\mathbb{R}}
\newcommand{\N}{\mathbb{N}}
\newcommand{\Z}{\mathbb{Z}}
\newcommand{\C}{\mathbb{C}}
\newcommand{\Q}{\mathbb{Q}}
\newcommand{\Set}[2]{\left\{ #1 \mid #2 \right\}}
\renewcommand{\Im}{\operatorname{Im}}
\renewcommand{\Re}{\operatorname{Re}}
\newcommand{\K}{\mathcal{K}}
\newcommand{\A}{\mathcal{A}}
\renewcommand{\O}{\mathcal{O}}

\begin{document}

\title[Decision problems for linear recurrences]{Decision problems for linear recurrences involving arbitrary real numbers}
\author[E. Neumann]{Eike Neumann}
\address{Max Planck Institute for Software Systems, Saarland Informatics Campus, Saarbrücken, Germany}
\email{eike@mpi-sws.org}
\thanks{This work was conducted while the author was affiliated with the Department of Computer Science, Oxford University, UK}

\begin{abstract}
    \noindent We study the decidability of the Skolem Problem, the Positivity Problem, and the Ultimate Positivity Problem for linear recurrences with real number initial values and real number coefficients in the bit-model of real computation.
    We show that for each problem there exists a correct partial algorithm which halts for all problem instances for which the answer is locally constant,
    thus establishing that all three problems are as close to decidable as one can expect them to be in this setting.
    We further show that the algorithms for the Positivity Problem and the Ultimate Positivity Problem halt on almost every instance with respect to the usual Lebesgue measure on Euclidean space. 
    In comparison, the analogous problems for exact rational or real algebraic coefficients are known to be decidable only for linear recurrences of fairly low order.
\end{abstract}

\maketitle

\section{Introduction}

A real linear recurrence sequence is a sequence $(u_k)_k$ of real numbers satisfying a linear recurrence relation of the form 
$u_{k + n} = c_1 u_{k + n - 1} + \dots + c_n u_k$.
The Skolem-Mahler-Lech theorem asserts that the zero set $\Set{k \in \N}{u_k = 0}$ of such a sequence is of a particularly simple form:
it is the union of a finite set $F$ and an empty or infinite set $I$ which is the union of a finite number of arithmetic progressions.
For given rational or integer coefficients $c_1,\dots, c_n$ and initial values $u_1,\dots,u_n$ the set $I$ can be effectively computed \cite{BerstelMignotte76}.
The same is not known of the set $F$.
It is well known \cite{Recurrences03} that the problem of computing the set $F$ is equivalent to the problem of deciding whether a given linear recurrence has a zero.
The latter problem, often referred to as the \emph{Skolem Problem} in the literature, has proven to be infamously difficult.
It is generally believed to be decidable for linear recurrences of any order, but known to be decidable only up to order four \cite{Mignotte1984, Vereshchagin}.
Two closely related decision problems of note are the Positivity Problem, which asks for given a linear recurrence to decide whether all of its terms are non-negative, and the Ultimate Positivity Problem, which asks for given a linear recurrence to decide whether all but finitely many of its terms are non-negative.
In \cite{OWPositivityLowOrder14} it was shown that both Positivity and Ultimate Positivity are decidable up to order five.
At the same time it was shown that a feasible algorithm for solving either Positivity or Ultimate Positivity at order six would entail major breakthroughs in the field of Diophantine approximation, 
making it highly unlikely for existing mathematical methods to allow for further progress to be made on these problems.
For linear recurrences with simple characteristic roots the Positivity Problem is known to be decidable up to order nine \cite{OWPositivitySimple14} and the Ultimate Positivity Problem is known to be decidable for all orders \cite{OWUltimateSimple14}.
These and related decision problems have numerous applications in theoretical computer science and beyond, see \cite{OWPositivityLowOrder14} and references therein.
See the survey \cite{OWSurvey15} for a more detailed historical overview and further related results.

In this paper we study the Skolem Problem, the Positivity Problem, and the Ultimate Positivity Problem for linear recurrences whose coefficients and initial values are arbitrary real numbers which are given as fast converging Cauchy sequences of rational numbers.
For topological reasons no non-trivial problem is decidable in this setting.
Instead one should ask if there exists a \emph{maximal partial algorithm} for deciding a given problem.
When such an algorithm is given a problem instance as input it either diverges or halts and outputs the correct answer for the decision problem.
It is required to halt on every problem instance for which the answer for the decision problem is locally constant.
We will call such problem instances \emph{robust instances}.
This ensures that its halting set contains the halting set of any correct partial algorithm for deciding the problem.

Besides being mathematically interesting in their own right, the real number versions of the Skolem Problem and its variants can be motivated by practical applications.
In most applications to engineering and the natural sciences the assumption that the input be given as an exact integer, rational number, or real algebraic number is quite unrealistic.
There it is usually more appropriate to assume that the inputs be known only approximately to finite accuracy, but with a known error bound.
This can be modelled by assuming that one is given a rational box which contains the problem instance of interest.
A maximal partial algorithm for deciding the real number version of a decision problem can be automatically translated to an algorithm which takes as input such a box, halts and outputs $1$ if the box is contained in the set of robust ``yes''-instances,
halts and outputs $0$ if the box is contained in the set of robust ``no''-instances,
halts and outputs $-1$ if the box contains both robust ``no''- and robust ``yes''-instances,
and diverges in all other cases.
Such an algorithm need not exist when the problem is decidable on rational inputs:
There exist sets $A \subseteq \R$ such that $A \cap \Q$ is decidable but $A$ is not maximally partially decidable.
One can for instance take the union of any decidable subset of $\Q$ with a singleton $\{x\}$ that is not co-c.e.~closed.
It is of course conversely true that there exist sets $A \subseteq \R$ that are maximally partially decidable such that $A \cap \Q$ is not decidable.
One can take for instance any undecidable co-c.e.~set of integers.

We will show that the real number versions of the Skolem Problem, the Positivity Problem, and the Ultimate Positivity Problem are maximally partially decidable for linear recurrences of any order.
Thus, the real number versions of all three problems are, in a sense, as close to decidable as one can expect them to be.
That this is achievable for real linear recurrences despite the hardness results for rational ones is perhaps not too surprising.
The existing decidability proofs for low orders only fail to generalise to higher orders due to the presence of ``critical'' problem instances where the dominant part of the exponential polynomial solution does not admit an exponential lower bound.
Since this situation is unstable under small perturbations one will not be obliged to decide the problems in these critical instances for real number inputs, thus avoiding the aforementioned hardness results.

Indeed, our proof consists almost entirely of translating the known decision methods for low-orders to the real number setting and noting that these already suffice to establish maximal partial decidability for all orders.
The proof is considerably more elementary than its counterparts for integer coefficients.
The use of Baker's theorem and similar deep results from analytic number theory can be entirely avoided.
On the other hand new problems appear in the real number setting that are absent from the discrete setting.
The study of the asymptotic behaviour of a linear recurrence relies heavily on the study of its exponential polynomial solution.
It is easy to see that in the real number setting the exponential polynomial solution is not in general computable from the linear recurrence.
This is where new ideas are required.
It is relatively easy to see that one can still compute those coefficients of the exponential polynomial solution which belong to simple characteristic roots.
This will suffice to computably recognise all robust instances of the Skolem Problem and the Positivity Problem,
and all robust ``yes''-instances of the Ultimate Positivity Problem.
For the Ultimate Positivity Problem there exist robust ``no''-instances whose dominant characteristic roots are not simple, which implies that the corresponding coefficients in the exponential polynomial solution do not depend continuously on the input.
These instances are by far the most difficult ones to handle.
They will be treated by reduction to the first-order theory of the reals.
Recall that the first-order theory of the reals is the first-order theory of the structure 
$\langle  
\R, 0, 1, +, \times, -, >, =
\rangle$.
By the Tarski-Seidenberg theorem \cite[Theorem 2.77]{BasuPollackRoy} this theory is decidable.
The main ideas in this case are best illustrated with the help of a simple example:

\begin{exa}\label{Example: proof idea ultimate positivity}
    Consider the linear recurrence 
    $u_1 = \pi$, $u_2 = 2\pi$, $u_3 = \pi$, 
    $u_{k + 1} = 3u_k - 3u_{k - 1} + u_{k - 2}$.
    Its characteristic polynomial is $P(x) = x^3 - 3x^2 + 3x - 1 = (x - 1)^3$ 
    and
    its exponential polynomial solution is 
    $u_k = \pi + k\pi - k(k - 1)\pi$.
    It is hence a ``no''-instance of the Ultimate Positivity Problem.
    Let us show how we can verify this computationally when the coefficients and initial values are given as sequences of approximations.
    This is not completely straightforward since, as mentioned earlier, the exponential polynomial solution does not depend continuously on the input. 

    Choose a small rational number $\varepsilon > 0$.
    We can compute the roots of the characteristic polynomial to error $\varepsilon$ to verify that all complex roots 
    are contained in the open disk $B(1, \varepsilon)$ of radius $\varepsilon > 0$ centred at $1$.
    We can numerically count the roots in this disk with multiplicity, to find that there are three roots counted with multiplicity.
    Therefore there are only finitely many possibilities for the configuration of these roots:
    \begin{enumerate}
        \item $\mathcal{R}_1$: There is one real root $\rho$ with multiplicity $3$.
        \item $\mathcal{R}_2$: There is a real root $\rho_{0}$ with multiplicity $2$ and a real root $\rho_1$ with multiplicity $1$ and $\rho_0 > \rho_1$.
        \item $\mathcal{R}_3$: There is a real root $\rho_0$ with multiplicity $1$ and a real root $\rho_1$ with multiplicity $2$ and $\rho_0 > \rho_1$.
        \item $\mathcal{R}_4$: There are three simple real roots $\rho_0, \rho_1, \rho_2$ with $\rho_0 > \rho_1 > \rho_2$.
        \item $\mathcal{R}_5$: There is one simple real root $\rho$ and two complex conjugate roots $\lambda$, $\bar{\lambda}$.
    \end{enumerate}
    Call $\mathcal{R}_1,\dots,\mathcal{R}_5$ the \emph{possible root configurations}.
    To each root configuration we can assign a \emph{characteristic polynomial},
    for instance 
    $P_1(x) = (x - \rho)^3$
    and
    $P_5(x) = (x - \rho) (x - \lambda) (x - \bar{\lambda})$.
    This gives rise to an \emph{associated linear recurrence}.
    For instance, the linear recurrence associated with $\mathcal{R}_1$ is 
    \[
        u_{k + 1} = 3\rho u_k - 3 \rho^2 u_{k - 1} + \rho^3 u_{k - 2}
    \]
    and the linear recurrence associated with $\mathcal{R}_5$ is 
    \[
        u_{k + 1} = (\rho + \lambda + \bar{\lambda}) u_k - (\lambda\rho + \bar{\lambda}\rho + \lambda\bar{\lambda}) u_{k - 1} + \rho\lambda\bar{\lambda}
    \]
    We can symbolically compute the exponential polynomial solutions of each of these linear recurrences.

    Each possible root configuration $\mathcal{R}_j$ can be assigned a \emph{domain} 
    $D_{\mathcal{R}_j}$ which is the set of all ``valid'' assignments to the variables that occur in the root configuration.
    For instance,
    \[
        D_{\mathcal{R}_2} = \Set{(\rho_0, \rho_1) \in (1 - \varepsilon, 1 + \varepsilon)^2}{\rho_0 > \rho_1}
    \]
    and 
    \[
        D_{\mathcal{R}_5} = (1 - \varepsilon, 1 + \varepsilon) \times (B(1,\varepsilon) \cap \mathbb{H})
    \]
    where $\mathbb{H} = \Set{z \in \C}{\Im z > 0}$.
    Note that up to identifying $\C$ with $\R^2$ each domain is a definable set in the first-order theory of the reals.
    We can substitute a point in the domain of a possible root configuration $\mathcal{R}$ for the variables of $\mathcal{R}$
    to obtain a linear recurrence, which we call the \emph{associated linear recurrence} of that point.
    Any such linear recurrence is a ``small perturbation'' of our original linear recurrence, enriched with additional information about the algebraic multiplicities of its characteristic roots.
    In particular our original linear recurrence can be recovered as the associated linear recurrence of some point in the domain of some possible root configuration - in this concrete case it is the linear recurrence associated with $1 \in D_{\mathcal{R}_1}$.

    To establish that our instance is not ultimately positive it would hence suffice to show that for all points in the domain of all possible root configurations the coefficient of the dominant real term in the exponential polynomial solution of the associated linear recurrence is negative.
    This is a well-known result about linear recurrences. 
    See Lemma \ref{Lemma: behaviour of non-positive dominant part} below for a formal statement.

    Let us carry this out for the root configuration $\mathcal{R}_2$.
    The dominant term in the exponential polynomial solution of the associated linear recurrence is $k \rho_0^{k - 1}$.
    An explicit symbolic calculation shows that its coefficient is equal to 
    \[
        \frac{u_3 - (\rho_0 + \rho_1) u_2 + \rho_0\rho_1 u_1}{\rho_0 - \rho_1}.
    \]

    We can compute rational approximations $v_1, v_2, v_3$ of the initial values to error $\varepsilon$.
    It then suffices to show that the following sentence holds true:
    \begin{align*}
        &\forall \rho_0. \forall \rho_1. \forall u_1. \forall u_2. \forall u_3.\\
        &
            |\rho_0 - 1| < \varepsilon 
            \land |\rho_1 - 1| < \varepsilon
            \land \rho_0 > \rho_1
            \land |u_1 - v_1| < \varepsilon
            \land |u_2 - v_2| < \varepsilon
            \land |u_3 - v_3| < \varepsilon\\
            &\rightarrow
            \frac{u_3 - (\rho_0 + \rho_1) u_2 + \rho_0\rho_1 u_1}{\rho_0 - \rho_1} < 0
    \end{align*}
    This sentence can be formulated in the first-order theory of the reals.
    Its truth is hence decidable by the Tarski-Seidenberg theorem.
    It is clear that the sentence is indeed true for sufficiently small $\varepsilon > 0$.
    This yields a semi-decision procedure for showing that for all points in $D_{\mathcal{R}_2}$ the coefficient of the dominant real term in the exponential polynomial solution of the associated linear recurrence is negative. 

    A similar reduction to the first-order theory of the reals can be carried out for the remaining root configurations 
    $\mathcal{R}_1,\mathcal{R}_3,\dots,\mathcal{R}_5$
    to show that the given linear recurrence is a robust ``no''-instance of Ultimate Positivity.

    Of course, we have merely verified a particular sufficient condition which happened to hold true for this specific instance.
    The full algorithm will be slightly more involved, but it will follow the same ideas.
\end{exa}

\section{Decision Problems for continuous data}

We work within the framework of represented spaces as introduced by Kreitz and Weihrauch \cite{KreitzWeihrauch} for countably based spaces and extended by Schr\"oder \cite{SchroederPhD, SchroederAdmissibility} to quotients of countably based spaces.
See \cite{WeihrauchBook,BrattkaPresser,PaulyRepresented} for introductions to this approach to computable analysis at varying levels of abstraction. 
It will suffice for our purpose to work with admissibly represented countably based $T_0$ spaces.
We will mainly use the notation from \cite{PaulyRepresented}.
In particular for a represented space $X$ we denote by $\K(X)$ the represented space of compact sets with the sequentialisation of the upper Vietoris topology, by $\O(X)$ the space of open sets with the Scott topology, and by $\A(x)$ the space of closed sets where a closed set $A$ is identified with its complement $A^C \in \O(X)$.

For the benefit of readers unfamiliar with computable analysis we will briefly recall the very basic ideas for computing on the space $\R^n$.

A rational interval is a closed interval with rational endpoints.
A rational box is a finite product of rational intervals. 
A point $x \in \R^n$ can be represented by a sequence $(B_j)_j$ of rational boxes such that each box contains $x$ and $\bigcap_{j \in \N} B_j = \{x\}$.
Any such sequence is called a \emph{name} for $x$.
The point $x$ is called computable if it has a computable name, \textit{i.e.}, there exists an algorithm which on input $j \in \N$ outputs a box $B_j$ such that the resulting infinite sequence $(B_j)_j$ is a name for $x$.
A function $f \colon \R^n \to \R^m$ is computable if there exists an algorithm which takes as input\footnote{An infinite input sequence can for instance be implemented as an infinite stream that is written on a special input tape or as an oracle.} a name
$(B_j)_j$ of a point $x$ and a natural number $k \in \N$ and outputs a rational box $C_k$, 
such that for every fixed name $(B_j)_j$ the resulting infinite sequence $(C_k)_k$ is a name for $f(x)$.
It is easy to see that any computable function is necessarily continuous with respect to the usual topology on $\R^n$.

One may object that this notion of computability models an unrealistic situation in which one has arbitrarily good approximations to a real vector available.
However, it is easy to prove that $f \colon \R^n \to \R^m$ is computable if and only if there exists an algorithm which takes as input a rational box $B$ and a positive rational number $\varepsilon > 0$, and returns as output a finite list of boxes $C_1,\dots,C_s$ such that each $C_j$ has width at most $\varepsilon$, each $C_j$ intersects the range $f(B)$, and $C_1 \cup \dots \cup C_s \supseteq f(B)$.
Moreover, such an algorithm can be effectively computed from an algorithm which computes $f$ in the above sense and vice versa.
Thus, this notion of computability captures precisely the idea that one can compute arbitrarily good information on $f(x)$ when $x$ is given to finite accuracy.

Let us now discuss decision problems in this context.
With any subset $A \subseteq \R^n$ one can associate a decision problem: 
given $x \in \R^n$ as input halt in finite time and output $1$ if and only if $x \in A$ or output $0$ if $x \notin A$.
It is easy to see that for continuity reasons the only subsets of $\R^n$ that are decidable in this sense are the empty set and $\R^n$ itself.

The next best thing one can hope for is to find an algorithm which decides the problem in as many points as possible.
This is somewhat of a folklore idea in computable analysis, but there does not appear to be an established standard terminology in the literature.
Let $X$ be a represented space.
A \emph{partial\footnote{The term ``partial algorithm'' is used here in the traditional sense of theoretical computer science. 
              In breach with the usual convention in computable analysis the algorithm's behaviour is constrained on the entire space.
              A computable analyst may hence prefer to view such an algorithm as a total algorithm with values in Kleene space $\mathbb{K} = \{0,1,\bot\}$.} 
algorithm for deciding} a set $A \subseteq X$ is an algorithm which takes as input a name of a point $x \in X$ and either diverges or halts in finite time and outputs $0$ or $1$.
We require that such an algorithm be extensional, \text{i.e.}, that its termination and output on termination depend only on the point $x$ but not on the choice of name\footnote{In the case of countably based spaces this is an inessential restriction as these admit open representations.}.
We further require that it be \emph{correct}, \text{i.e.}, that it halt and return $1$ only when $x \in A$ and that it halt and return $0$ only when $x \notin A$.
Its \emph{halting set} is the set of points for which it halts.
This set is well-defined by the extensionality assumption.
A partial algorithm for deciding $A$ is called \emph{maximal} if its halting set contains the halting set of all other partial algorithms for deciding $A$. 
We call $A$ \emph{maximally partially decidable} if there exists a maximal partial algorithm for deciding $A$.

\begin{prop}\label{Proposition: characterisation of maximal partial decidability via points of continuity}
    Let $X$ be an admissibly represented countably based $T_0$ space.
    A partial algorithm for deciding $A$ is maximal if and only if its halting set is equal to the set of points of continuity of the characteristic function 
    $\chi_A \colon X \to \{0,1\}$, where $\{0,1\}$ carries the discrete topology.
\end{prop}
\begin{proof}
    It is easy to see that the halting set of an algorithm for deciding $A$ must be contained in the set of points of continuity of the characteristic function.
    Conversely, if $x$ is a point of continuity of $\chi_A$ then there exists a basic open set $B$ which contains $x$ such that $\chi_A$ is constant on $B$.
    There exists an algorithm which halts on $B$ and outputs the constant value of $\chi_A$.
    It follows that any maximal partial algorithm for deciding $A$ must contain $x$ in its halting set.
\end{proof}

In other words, a maximal partial algorithm for deciding $A$ is an algorithm that takes as input $x \in X$, halts and outputs $1$ if $x$ is contained in the interior of $A$, halts and outputs $0$ if $x$ is contained in the interior of the complement of $A$, and diverges if $x$ is contained in the boundary of $A$.
In particular, every set is maximally partially decidable relative to some oracle, and if $X$ is a discrete space then $A \subseteq X$ is maximally partially decidable if and only if it is decidable.
Therefore maximal partial decidability seems to be, in some sense, a more appropriate generalisation of decidability over $\N$ than ``naive'' decidability.
One should however bear in mind that this is a relative notion:
The set of rational numbers $\Q$, say, is a maximally partially decidable subset of $\R$.
A maximal partial correct algorithm is given by the algorithm that never halts.
Once maximal partial decidability of a problem is established one is hence naturally lead to the study of the ``absolute size'' of the halting set. 

It will be convenient to introduce the following terminology for decision problems:
Let $A \subseteq X$ be a set.
Call any point $x \in X$ an \emph{instance of (the decision problem associated with)} $A$.
If $x \in A$ then $x$ is called a \emph{``yes''-instance}.
If $x \notin A$ then $x$ is called a \emph{``no''-instance}.
If $x$ is a point of continuity of the characteristic function $\chi_A$, \textit{i.e.}, if $x$ is not contained in the boundary of $A$,
then $x$ is called a \emph{robust instance} of $A$.

Finally, the following convention will be very useful:
We say that a property $P$ holds true for a point $p \in X$ \emph{up to an arbitrarily small perturbation} if for every open set $U \in \O(X)$ which contains $x$ there exists $y \in U$ such that $P$ holds true for $y$.

As mentioned in the introduction, maximal partial decidability is equivalent to ``almost deciding'' a trichotomy when the input is known only to finite accuracy.
We formulate this only for the case of $\R^n$ but it generalises easily to all locally compact spaces.

\begin{prop}\label{Proposition: maximal partial decidability and decidability with finite accuracy}
    Let $A \subseteq \R^n$ be a set.
    Then $A$ is maximally partially decidable if and only if there exists an algorithm 
    which takes as input a rational box $B$,
    halts and outputs $1$ if $B$ is contained in the set of robust ``yes''-instances of $A$,
    halts and outputs $0$ if $B$ is contained in the set of robust ``no''-instances of $A$,
    halts and outputs $-1$ if $B$ contains robust ``yes''-instances as well as robust ``no''-instances,
    and diverges in all other cases.

    Moreover, such an algorithm can be effectively computed from a maximal partial algorithm for deciding $A$ and vice versa.
\end{prop}
\begin{proof}
    It is obvious that the existence of such an algorithm implies maximal partial decidability.

    Assume that $A$ is maximally partially decidable.
    Then the set of robust ``yes''-instances is computable as an element of the represented space $\O(\R^n)$ of open sets with the Scott topology.
    The same holds true for the set of ``no''-instances.

    An encoding of a rational box $B$ can be effectively translated to a name of the same box as an element of the space 
    $\K(\R^n)$ of compact subsets with the sequentialisation of the upper Vietoris topology.
    One can therefore semi-decidable if all instances contained in $B$ are robust ``yes''-instances or if all instances contained in $B$ are robust ``no''-instances.

    We can also effectively compute a name of the interior of $B$ as an element of the space $\O(\R^n)$ of open sets with the Scott topology.
    Now, if $B$ contains a robust ``yes''-instance then the interior $B^{\circ}$ of $B$ contains a robust ``yes''-instance.
    We can thus semi-decide if $B$ contains a robust ``yes''-instance by computing the intersection of $B^{\circ}$ with the set of robust ``yes''-instances as an element of $\O(\R^n)$ and semi-deciding if the result is non-empty.
    By symmetry the same is true for robust ``no''-instances.
    The claim follows.
\end{proof}

Throughout this section we have focussed our attention on countably based spaces.
This will be sufficient for the purpose of this paper.
It should be pointed out however, that Proposition 
\ref{Proposition: characterisation of maximal partial decidability via points of continuity}
may fail for non-countably-based spaces.
To see this, let 
\[
    \ell^2 = 
        \Set{(x_n)_n \in \R^{\N}}{\sum_{n \in \N} x_n^2 < +\infty}
\]
be the computable metric space of square-summable real sequences with the metric induced by the usual inner product
$\langle\cdot,\cdot\rangle$.
Let $\left(\ell^2\right)' \subseteq \R^{\ell^2}$ denote the admissibly represented space of linear functionals on $\ell^2$,
with the representation inherited from the exponential $\R^{\ell^2}$ in the category of represented spaces.
As usual, the space $\left(\ell^2\right)'$ can be identified with $\ell^2$ (with a weaker topology) by virtue of the self-duality of Hilbert spaces.
In particular, the inner product $\langle \cdot, \cdot \rangle$ can be defined on $\left(\ell^2\right)'$.
It is shown in \cite{BrattkaSchroeder05} that 
the space $\ell^2$ is computably isomorphic to the space 
$\Set{(x_n)_n \in \R^{\N}}{x_0 = \sum_{n \geq 1} x_n^2}$.
It is further shown that 
the space $\left(\ell^2\right)'$ is computably isomorphic to the quotient space 
\[ 
    \Set{(x_n)_n \in \R^{\N}}{x_0 \geq \sum_{n \geq 1} x_n^2}/\sim
\]
where $(x_n)_n \sim (y_n)_n$ if and only if $(x_n)_{n \geq 1} = (y_n)_{n \geq 1}$.
In other words, a point $x \in \ell^2$ can be represented by the sequence $(x_n)_n$ and its $\ell^2$-norm,
while a point $x \in \left(\ell^2\right)'$ can be represented by the sequence $(x_n)_n$ and some upper bound on its $\ell^2$-norm.

Now, let $p \in \left(\ell^2\right)'$ be a computable point with uncomputable $\ell^2$-norm (see \textit{e.g.}~\cite[Theorem 5.9]{NeumannBGK} for a construction of a such a point).
Consider the hyperplane 
\[ 
    H = \Set{x \in \left(\ell^2\right)'}{\langle p, x \rangle = 0}.
\]
Then the complement of $H$ is open and dense but contains no non-empty semi-decidable subset.
To see the last claim, assume that there exists an algorithm whose halting set is a non-empty subset $U$ of the complement of $H$.
Identify $\left(\ell^2\right)'$ with 
\[ 
    \Set{(x_n)_n \in \R^{\N}}{x_0 \geq \sum_{n \geq 1} x_n^2}/\sim
\]
as above.
Let $x = (x_0,x_1,\dots)$ be a point in $U$.
Then there exists an integer $m$ such that the algorithm halts 
on all points of the form 
$(N, x_1,\dots,x_m, 0, 0, \dots)$
with $N \geq x_1^2 + \dots + x_m^2$.
From this it follows that 
for every $N > x_1^2 + \dots + x_m^2$ there exists a positive integer $\nu(N)$ such that the algorithm halts on all sequences in $\left(\ell^2\right)'$ that start with
$N, x_1,\dots, x_m$, followed by $\nu(N)$ zeroes.
In particular the algorithm halts on all sequences
\[ 
    s(\alpha) = \left(N, x_1,\dots, x_m,\underbrace{0,\dots,0}_{\nu(N)\text{ times}}, \alpha \cdot p_{m + \nu(N) + 1}, \alpha \cdot p_{m + \nu(N) + 2},\dots\right)
\]
with 
$x_1^2 + \dots + x_m^2 + \alpha^2 \sum_{n > m + \nu(N)} p_n^2 \leq N$.
By solving the equation $\langle p, s(\alpha) \rangle = 0$ for $\alpha$
and using that the halting set is assumed not to intersect $H$, 
we find that 
\[
    \sum_{n > m + \nu(N)} p_n^2 < \frac{\left(p_1 x_1 + \dots + p_m x_m\right)^2}{N - \left(x_1^2 + \dots + x_m^2\right)}.
\]
Since $N$ may be chosen arbitrarily large, this yields an algorithm for computing the norm of $p$, contradicting our initial assumption.

It follows that $H$ is maximally partially decided by the algorithm that never halts.
The halting set of this algorithm is clearly much smaller than the points of continuity of the characteristic function of $H$.

This suggests that a reasonable alternative approach to the study of the ``partial decidability'' of subsets of admissibly represented spaces
may be obtained by taking the characterisation given in Proposition 
\ref{Proposition: characterisation of maximal partial decidability via points of continuity}
as the definition directly.
Letting $\mathbb{K} = \{0, 1, \bot\}$
be endowed with the topology generated by the sets $\{0\}$ and $\{1\}$,
one can study the computability of the continuous function
$
    \widetilde{\chi}_A \colon X \to \mathbb{K}
$
which sends the interior of $A$ to $1$,
the complement of the closure of $A$ to $0$,
and the boundary of $A$ to $\bot$.
For countably based $X$ one recovers the definition of maximal partial decidability via Proposition \ref{Proposition: characterisation of maximal partial decidability via points of continuity}.
Since the hyperplane $H$ above is maximally partially decided by the algorithm that never halts,
the function $\widetilde{\chi}_{H}$ is ``nowhere computable'' in the sense that
its restriction to every open subset of $\left(\ell^2\right)'$ is uncomputable.
The function $\widetilde{\chi}_{A}$ arises quite naturally as the continuous coreflection of the characteristic function 
$\chi_A \colon X \to \mathbb{K}$ 
in the sense of Escard\'o \cite[Proposition 2.6.1]{EscardoInjective}
-- an idea that already goes back to Scott \cite{ScottLattices}.
It is also the best continuous approximation to $\chi_A$ in the very strong sense of \cite{NeumannPhD}.

\section{Linear Recurrences}

A real linear recurrence sequence is a sequence $(u_k)_k$ such that there exists a positive integer $n \in \N$ 
and real numbers $c_1,\dots,c_n \in \R$ such that 
\begin{equation}\label{eq: linear recurrence} 
    u_{k} = c_1 u_{k - 1} + c_2 u_{k - 2} + \dots + c_n u_{k - n}
\end{equation}
for all $k > n$.
This sequence can hence be encoded by the vector 
\[
    (c,u) = (c_1,\dots,c_n, u_1,\dots,u_n) \in \R^{2n}.
\]
Note that this encoding is not unique.
The \emph{order} of a linear recurrence sequence is the smallest possible $n$ such that $(u_k)_k$ satisfies a relation of the above form. 

From a computational point of view it is important to treat different encodings of the same sequence as different problem instances.
This will be illustrated in Example \ref{Example: motivation of basic definitions} below.
We define a \emph{linear recurrence} to be a vector $(c,u) \in \R^{2n}$ for some $n \in \N$.
The number $n$ is called the \emph{order} of $(c,u)$.
Note that the sequence $(u_k)_k$ generated by the linear recurrence $(c,u)$ could satisfy a linear recurrence relation of strictly lower order.
In other words, the order of the linear recurrence is in general not the same as the order of the linear recurrence sequence $(u_k)_k$ it generates.
The space of all linear recurrences is identified with the represented space $\coprod_{n \in \N} \R^{2n}$.

The \emph{companion matrix} of a linear recurrence $(c,u)$ is the matrix 
\[
 \begin{pmatrix}
    c_1   & c_2 & \dots     & c_{n - 1} & c_n       \\
    1     & 0   & \dots     & 0         &  0        \\ 
    0     & 1   &  \dots         & 0         &  0        \\ 
    \vdots & \vdots   &  \ddots   & \vdots & \vdots    \\
    0     & 0   & \dots     & 1         &  0 
 \end{pmatrix}   
\]

The \emph{characteristic polynomial} of $(c,u)$ is the polynomial
\[
    P(x) = x^n - c_1x^{n - 1} - \dots - c_n,
\]
\textit{i.e.}, up to sign, the characteristic polynomial of its companion matrix.
The complex roots of the characteristic polynomial are called the \emph{characteristic roots} of $(c,u)$.

The \emph{exponential polynomial solution} plays a crucial role in the study of the asymptotic behaviour of linear recurrences.
Its definition is a bit subtle in our context.
A \emph{formal complex polynomial} is a vector $Q = (a_0,\dots,a_m) \in \C^m$, which we also write as $Q(x) = a_0 + \dots + a_m x^m$.
The \emph{formal degree} of $Q$ is the number $m$.
We do not assume here that $a_m \neq 0$, hence the name ``formal degree''.

A \emph{formal exponential polynomial} is a function $f \colon \N \to \C$ of the form 
\[
    f(k) = P_1(k) \lambda_1^k + \dots + P_s(k) \lambda_s^k
\]
where the $P_j$'s are formal polynomials and the $\lambda_j$'s are distinct complex numbers.
A formal exponential polynomial is assumed to be encoded as a vector in $\C^{d_1}\times\dots\times\C^{d_s}\times\C^s$,
where $d_1,\dots,d_s$ are the formal degrees of $P_1,\dots,P_s$.
The space of formal exponential polynomials is the co-product over all spaces of this form.
A formal exponential polynomial is \emph{formally real-valued} if for every $\lambda_j \in \C \setminus \R$ there exists 
an index $l$ with $\lambda_l = \bar{\lambda}_j$ and $P_l = \bar{P}_j$.
Note that this is stronger than to require for the function $f(k)$ to be real-valued.
The formal exponential polynomial $1^k + 0 \cdot i^k$ is real-valued but not formally real-valued.

The \emph{exponential polynomial solution} of a linear recurrence $(c,u)$ is the unique formal exponential polynomial
$
    f(k) = P_1(k) \lambda_1^k + \dots + P_s(k) \lambda_s^k
$
satisfying $u_k = f(k)$ for all $k$, 
where $\lambda_1,\dots,\lambda_s$ are the distinct characteristic roots of the characteristic polynomial $P$ of $(c,u)$ and the 
formal degree of the formal polynomial $P_j$ is the multiplicity of $\lambda_j$ as a root of $P$.
Its existence and uniqueness follow from existence and uniqueness of the Jordan normal form of the companion matrix of $(c,u)$.
Note that the encoding of $f$ as a vector in $\C^{d_1}\times\dots\times\C^{d_s}\times\C^s$ is unique only up to permutation.
The exponential polynomial solution is clearly formally real valued.
Conversely, every formally real-valued exponential polynomial in $\C^{d_1}\times\dots\times\C^{d_s}\times\C^s$ is the exponential 
polynomial solution of a unique linear recurrence $(c,u) \in \R^{2(d_1 + \dots + d_s)}$.

The \emph{dominant characteristic roots} are the roots of the characteristic polynomial of maximal modulus.
Let $d$ be the greatest multiplicity among dominant characteristic roots.
The \emph{formal dominant part} of the exponential polynomial solution is the formal exponential polynomial
\[
    a_{j_1} k^d \lambda_{j_1}^k + \dots + a_{j_t} k^d \lambda_{j_t}^k
\]
where $\lambda_{j_1},\dots,\lambda_{j_t}$ are the dominant characteristic roots of multiplicity $d$ 
and $a_{j_1},\dots,a_{j_t}$ are the coefficients of $x^d$ in the formal polynomials $P_{j_1},\dots,P_{j_t}$.
Note that it is possible for any of the $a_{j_l}$'s to vanish, so that the formal dominant part could vanish everywhere as a function.

We can now introduce the problems of interest more formally.
The Positivity Problem is the decision problem for the set 
\[
    \Set{(c,u) \in \coprod_{n \in \N} \R^{2n}}{u_k \geq 0 \text{ for all }k \in \N}.
\]
The Ultimate Positivity Problem is the decision problem for the set 
\[
    \Set{(c,u) \in \coprod_{n \in \N} \R^{2n}}{\text{there exists } K \in \N \text{ such that } u_k \geq 0 \text{ for all }k \geq K}.
\]
The Skolem Problem is the decision problem for the set 
\[
    \Set{(c,u) \in \coprod_{n \in \N} \R^{2n}}{u_k = 0 \text{ for some }k \in \N}.
\]

Here we have chosen to treat different encodings of the same linear recurrence sequence as different problem instances.
Since the above decision problems pertain to extensional properties of the sequences themselves, 
it may seem more natural to consider them on the quotient of the space 
$\coprod_{n \in \N} \R^{2n}$
under the identification of linear recurrences that encode the same sequence.
Such identifications are commonly performed in the computable analysis literature in analogous situations,
such as in the definition of the space of polynomials \cite{CallardHoyrup20,Hoyrup20,OvertChoice20} or the space of analytic functions \cite{PaulySteinberg17}.

\begin{exa}\label{Example: motivation of basic definitions}
    The following example should illustrate why one should discuss decidability on the level of encodings rather than on the level of sequences,
    why one should treat the exponential polynomial solution as a formal exponential polynomial,
    and why the asymptotic behaviour of the formal dominant part is more relevant than the asymptotic behaviour of the exponential polynomial as a function.

    Consider the sequence $(u_k)_k$ with $u_k = 1$ for all $k$.
    This sequence is strictly positive and therefore a-fortiori ultimately positive and without zeroes.

    The sequence can be viewed as a first-order linear recurrence satisfying the linear recurrence relation $u_k = u_{k - 1}$.
    It can therefore be encoded as the vector $(1,1) \in \R^2$.
    Its characteristic polynomial is $P(x) = x - 1$,
    its only characteristic root is $1$ with multiplicity $1$, 
    and its exponential polynomial solution is 
    $f(k) = 1 \cdot 1^k$.
    The formal exponential polynomial $f$ is equal to its formal dominant part.  
    It is easy to see that this linear recurrence remains strictly positive under small perturbations of the coefficients and initial values.
    It is therefore a robust ``yes''-instance of Positivity and Ultimate Positivity and a robust ``no''-instance of the Skolem Problem.

    The same sequence can also be viewed as a second-order linear recurrence satisfying the linear recurrence relation 
    $u_k = 2u_{k - 1} - u_{k - 2}$.
    This yields a different encoding as the vector $(2,-1,1,1) \in \R^4$.
    The characteristic polynomial of this linear recurrence is 
    $Q(x) = x^2 - 2x + 1 = (x - 1)^2$,
    its only characteristic root is $1$ with multiplicity $2$,
    and its exponential polynomial solution is 
    $g(k) = (0\cdot k + 1) \cdot 1^k$.
    The formal dominant part of $g$ is $0 \cdot k \cdot 1^k$.
    Note that the formal exponential polynomial $g$ is equal as a function to the formal exponential polynomial $f$ above. 
    For $N \in \N$, consider the linear recurrence encoded by 
    $(2,-1, 1 - \tfrac{1}{N}, 1 - \tfrac{2}{N})$.
    The exponential polynomial solution becomes 
    $(-\tfrac{1}{N} k + 1) \cdot 1^k$.
    For $k = N$ this new linear recurrence has a zero and for $k > N$ it becomes strictly negative.
    Since $N$ can be chosen arbitrarily large, we obtain arbitrarily small perturbations of the initial values such that 
    the resulting linear recurrence is a ``no''-instance of Positivity and Ultimate Positivity and a ``yes''-instance of the Skolem Problem.
    Therefore, this linear recurrence is not a robust instance of any of these problems. 
\end{exa}

It is easy to extend the idea of Example \ref{Example: motivation of basic definitions} to show that the problems of interest become trivial on the represented space $\coprod_{n \in \N} \R^{2n}/\sim$, where $\sim$ is the relation that identifies linear recurrences which encode the same sequence,
in the sense that all three problems are maximally partially decided by the algorithm that never halts.
The same holds true if one is given a linear recurrence sequence $(u_k)_k$ as an element of the represented space $\R^{\N}$ together with a bound $n$ on its order.
Another sensible way of encoding a linear recurrence sequence $(u_k)_k$ is to provide a matrix $A$ and two vectors $v$ and $w$ such that $u_k = v A^k w$.
But this is easily seen to be equivalent to the encoding that we have chosen.
Finally, it also makes sense to encode a linear recurrence directly as a formal exponential polynomial.
This is clearly a strictly stronger representation than the one we have chosen.
It is relatively straightforward to show based on the proofs given in this paper that the Skolem Problem, the Positivity Problem, and the Ultimate Positivity Problem for exponential polynomials are maximally partially decidable.
This is in fact much easier than the analogous problem for linear recurrences given by a vector of coefficients and a vector of initial values.

We will make frequent use of the following well-known result on linear recurrences.
For a proof see \textit{e.g.} \cite[Theorem 2]{BellGerhold07}. 

\begin{lem}\label{Lemma: behaviour of non-positive dominant part}
    Let $(c,u)$ be a linear recurrence which is not identically zero.
    Assume that $(c,u)$ has a dominant characteristic root which is not a positive real number.
    Let $\lambda_1,\dots,\lambda_n \in \C \setminus [0,+\infty)$
    be the non-positive dominant characteristic roots of $(c,u)$.
    Let $P_1,\dots,P_n$ be their respective coefficients in the exponential polynomial solution.
    Then the exponential polynomial 
    $P_1(k) \lambda_1^k + \dots + P_n(k) \lambda_n^k$ 
    is either identically zero or admits infinitely many positive and infinitely many negative values.
\end{lem}

The next result says that a linear recurrence depends continuously on its exponential polynomial solution.
In other words, small perturbations of the exponential polynomial solution induce small perturbations of the linear recurrence.
This fact will be extremely useful.

\begin{prop}\label{Proposition: linear recurrence depends continuously on exponential polynomial solution}
    Let $n \in \N$. 
    Let $d_1,\dots,d_s \in \N$ with $d_1 + \dots + d_s = n$.
    There exists a surjective computable map 
    \[
        \psi \colon \subseteq \C^{d_1}\times\dots\times\C^{d_s} \times \C^s \to \R^{2n}
    \]
    which sends a vector that encodes a formally real-valued formal exponential polynomial $f$
    to the unique linear recurrence $(c,u)$ with $f(k) = u_k$ for all $k \in \N$.
\end{prop}
\begin{proof}
    Assume we are given as input a vector 
    $(P_1,\dots,P_s,(\lambda_1,\dots,\lambda_s)) \in \C^{d_1}\times\dots\times\C^{d_s} \times \C^s$,
    where the $P_j$'s are interpreted as formal polynomials,
    such that the exponential polynomial $f(k) = P_1(k) \lambda_1^k + \dots + P_s(k) \lambda_s^k$
    is formally real-valued for all $k$.

    To obtain the vector $u$ we can simply evaluate $f$ for $k = 1,\dots,n$.
    A priori this yields $u_1,\dots,u_n$ as complex numbers, but since these numbers
    are guaranteed to be real we can compute their real part to obtain the same 
    numbers as real numbers.

    To obtain the coefficients $c$ of the linear recurrence,
    we can effectively compute the vector of coefficients of the polynomial 
    $(x - \lambda_1)^{d_1} \cdot \dots \cdot (x - \lambda_s)^{d_s}$
    as a vector of complex numbers.
    Since all coefficients are guaranteed to be real we can compute the real part of each entry 
    to obtain $c$ as a real vector.
\end{proof}

Similarly, the coefficients of a linear recurrence depend continuously on its characteristic roots in the following sense:

\begin{prop}\label{Proposition: coefficients depend continuously on characteristic roots}
    Let $n \in \N$.
    Let $D \subseteq \C^n$ be the set of all complex vectors whose entries constitute the roots of some monic polynomial with real coefficients, counted with multiplicity.
    Let $h \colon D \to \R^n$ be the function that sends a complex vector $(\lambda_1,\dots,\lambda_n)$ to the unique vector
    $(c_1,\dots,c_n) \in \R^n$ such that 
    $(x - \lambda_1) \cdot \dots \cdot (x - \lambda_n) = x^n - c_1 x^{n - 1} - \dots - c_n$.
    Then the map $h$ is computable.
\end{prop}

\section{On the computability of the exponential polynomial solution}

The aim of this section is to establish that the coefficients in the exponential polynomial solution of any simple dominant characteristic root are computable. 
First we recall that the characteristic roots are computable: 

\begin{thmC}[{\cite{SpeckerZeroes}}]\label{Theorem: Specker Zeroes}
    There exists an algorithm which takes as input a complex vector 
    $(a_0,\dots,a_d) \in \coprod_{d \geq 1} \C^{d + 1}$ 
    with $a_d \neq 0$ and outputs a vector 
    $(\lambda_1,\dots,\lambda_d)\in \C^d$
    such that 
    $(\lambda_1,\dots,\lambda_d)$
    contains the roots of the polynomial 
    $a_0 + a_1 z + \dots + a_d z^d$,
    counted with multiplicity.
\end{thmC}

It is worth mentioning that the algorithm in Theorem \ref{Theorem: Specker Zeroes} is non-extensional: the output vector depends on the name of the input vector and not just on the vector itself.
This is not surprising, since it is well known that there is no continuous single-valued function which assigns to the coefficient vector of a complex polynomial its vector of complex roots.

\begin{lem}\label{Proposition: geometric multiplicity equal to one}
    Let $(c,u)$ be a linear recurrence.
    Let $\lambda$ be a characteristic root of $(c,u)$. 
    Then the geometric multiplicity of $\lambda$ in the companion matrix of $(c,u)$ is equal to $1$.
\end{lem}
\begin{proof}
    The companion matrix is 
    \[
        A =
        \begin{pmatrix}
            c_1 & c_2 & \dots & c_n \\ 
            1   & 0   & \dots & 0   \\
                & \ddots & \ddots & \vdots \\ 
               &   & 1      & 0
        \end{pmatrix}
    \]
    Hence 
    \[
        A - \lambda I =
        \begin{pmatrix}
            c_1 - \lambda & c_2 & \dots & c_n \\ 
            1   & -\lambda   &  & 0   \\
                & \ddots & \ddots &  \\ 
               &   & 1      & -\lambda
        \end{pmatrix}
    \]
    Let $(x_1,\dots,x_n)$ be an eigenvector of $A$.
    Then for all $j = 2,\dots,n$ we have the equation $x_{j - 1} = \lambda x_j$.
    Since any eigenvector is by definition non-zero it follows that $x_n \neq 0$. 
    It then follows that $(\lambda^{n - 1},\dots,\lambda,1)$ is an eigenvector
    and that every eigenvector is a multiple of this one.
    Hence the geometric multiplicity of $\lambda$ is equal to $1$.
\end{proof}

\begin{lem}\label{Lemma: computing partial Jordan normal form}
    Let $A \in \R^{n \times n}$.
    Let $\mu$ be an eigenvalue of $A$ with geometric multiplicity $1$ and algebraic multiplicity $m$.
    Let $v_1,v_2,\dots,v_{m}$ be a Jordan chain of length $m$ for $A$, \textit{i.e.}, 
    $v_1$ is an eigenvector with eigenvalue $\mu$ and $v_{i - 1} = Av_i - \mu v_i$ for $i > 1$.
    Then we can uniformly compute in $v_1, \dots, v_{m}$ and $A$ an invertible matrix $S \in \C^{n \times n}$
    such that 
    \[
        S A S^{-1} = 
            \begin{pmatrix}
                J & 0 \\
                0 & B
            \end{pmatrix}
    \]
    where $J$ is a Jordan block for $\mu$ of size $m$ and all generalised eigenvectors of $S A S^{-1}$
    for eigenvalues $\lambda$ of $A$ with $\lambda \neq \mu$ lie in the span of the standard unit vectors $e_{m + 1},\dots, e_n$.
\end{lem}
\begin{proof}
    The union of all generalised eigenspaces for any eigenvalue $\lambda$ of $A$ is uniformly computable from $\lambda$ as an element of $\A(\C^n)$.
    It follows that the union of all generalised eigenspaces for all $\lambda \neq \mu$ can be computed as an element of $\A(\C^n)$.
    By intersecting with the unit sphere we can compute the set of all normalised generalised eigenvectors for all eigenvalues $\lambda \neq \mu$ as an element of $\K(\C^n)$.
    Since linear independence is semi-decidable it follows that the linear span of the generalised eigenvectors which belong to eigenvalues other than $\mu$ is computable as an element of $\A(\C^n)$.
    Since this is a linear space whose dimension is known to be $n - m$ it follows from {\cite[Theorem 11]{BrattkaZieglerLA}} that we can compute a basis $b_1, \dots, b_{n - m}$ of this space\footnote{Theorem 11 in \cite{BrattkaZieglerLA} is formulated over $\R^n$ but uses only Hilbert space properties of the reals and therefore carries over to $\C^n$ without modification (cf.~the remark in the second paragraph of page 191 in \cite{BrattkaZieglerLA}).}.
    
    We can thus compute the matrix $S$ which sends $v_1,\dots,v_m$ to $e_1,\dots,e_m$ and $b_1,\dots,b_{n - m}$ to $e_{m + 1},\dots, e_n$.
    The result follows immediately.
\end{proof}

\begin{cor}\label{Corollary: computing coefficient of eigenvalue with known multiplicity}
    Let $(c,u)$ be a linear recurrence.
    Let $\mu$ be an eigenvalue of its companion matrix $A$ with algebraic multiplicity $m$.
    Then coefficients of the formal polynomial coefficient of $\mu^k$ in the exponential polynomial solution of $(c,u)$ are
    uniformly computable in $u$, $c$, and $m$
    as a vector in $\C^m$.
    Moreover, if $\mu$ is real, then the coefficients are computable as a vector in $\R^m$.
\end{cor}
\begin{proof}
    By Proposition \ref{Proposition: geometric multiplicity equal to one} the geometric multiplicity of $\mu$ is equal to $1$.
    By Lemma \ref{Lemma: computing partial Jordan normal form} we can thus compute a matrix $S$ such that 
    \[S A S^{-1} 
    =
    \begin{pmatrix}
        J & 0 \\
        0 & B
    \end{pmatrix}\]
    with $J$ being a Jordan block for $\mu$ of size $m$ and all further generalised eigenspaces lying in the span of the standard unit vectors 
    $e_{m + 1}, \dots, e_n$.

    Let $v = (0,\dots,0,1) S^{-1}$.
    Let $w = S u$.
    We claim that the term belonging to $\mu^k$ in the exponential polynomial solution of the linear recurrence is equal to 
    $(v_1,\dots,v_m)^T J^k (w_1,\dots, w_m)$.

    Since $e_1,\dots,e_m$ form a Jordan chain for the eigenvalue $\mu$ of $S A S^{-1}$
    and all remaining Jordan chains are by assumption contained in the span of $e_{m + 1},\dots,e_n$
    the matrix $S A S^{-1}$ can be put into Jordan normal form by 
    conjugation with a matrix
    \[
        T =
        \begin{pmatrix}
            I & 0 \\
            0 & M
        \end{pmatrix}
    \]
    where $I$ is the $m \times m$ identity matrix and $M$ is some invertible $(n - m) \times (n - m)$-matrix.
    The exponential polynomial solution of the linear recurrence $(c,u)$ is thus given by 
    \[
        v^T 
        T^{-1}
        \begin{pmatrix}
            J^k & 0 \\
            0 & C^k
        \end{pmatrix}
        T
        w,
    \]
    where $C$ is an $(n - m) \times (n - m)$-matrix in Jordan normal form whose eigenvalues are different from $\mu$.
    It follows that the term belonging to $\mu^k$ is indeed $(v_1,\dots,v_m)^T J^k (w_1,\dots, w_m)$.
    It is clear that this can be computed as a complex vector.
    If $\mu$ is guaranteed to be real we can compute the real part of each entry of the vector to obtain the same vector as a real vector.
\end{proof}

\begin{lem}\label{Lemma: computing largest real coefficient}
    There exists an algorithm which takes as input a vector 
    $(a_0,\dots,a_n) \in \R^{n + 1}$
    with $a_n \neq 0$
    and a rational number $\varepsilon > 0$
    and
    halts if and only if the polynomial 
    $P(z) = a_0 + \dots + a_n z^n$
    has a real root and the largest real root of $P$ is simple,
    and on halting returns a rational approximation of the largest real root of $P$ to accuracy $\varepsilon$.
\end{lem}
\begin{proof}[Proof Sketch]
    Compute a list of intervals $I_1,\dots, I_s$ of width at most $\varepsilon$ such that every real root of $P$ is contained in some interval $I_j$.
    Let $I_k$ be such that the left endpoint of $I_k$ is larger than the right endpoint of every interval $I_j$ with $j \neq k$.
    Test to accuracy $\varepsilon$ if $P$ changes its sign on the endpoints of $I_k$.
    If this is the case evaluate $P'$ on $I$ using interval arithmetic with precision $\varepsilon$.
    If the resulting interval does not contain zero halt and output the centre of $I_k$.
    If any of the above tests fail, rerun the algorithm with $\varepsilon/2$ instead of $\varepsilon$.
\end{proof}

\begin{lem}\label{Lemma: computing dominant characteristic real root and its coefficient}
    There exists an algorithm which takes as input a linear recurrence $(c,u)$ and a rational number $\varepsilon > 0$,
    halts if and only if the linear recurrence has a unique and simple dominant characteristic root $\rho$ which in addition is a positive real number, 
    and on halting outputs $\rho$ together with a rational approximation to error $\varepsilon$ of its coefficient $a \in \R$ in the exponential polynomial solution of $(c,u)$.
\end{lem}
\begin{proof}
    Given any linear recurrence $(c,u)$ we can compute a vector in $\C^n$ containing all the complex roots of its characteristic polynomial.
    We can then semi-decide if the vector contains a unique element $\mu$ of maximal modulus.
    Moreover, we can semi-decide if there exists a rational box $B$ which is symmetric about the real axis and contains $\mu$ and no other characteristic roots.
    This then establishes that $\mu$ is a simple root of the characteristic polynomial and a real number. 
    By computing $\Re \mu$ we obtain a name of $\mu$ as a real number. 
    We can then semi-decide if $\mu > 0$.
    By Corollary \ref{Corollary: computing coefficient of eigenvalue with known multiplicity} we can compute the coefficient $a$ of $\mu$ in the exponential polynomial solution.
\end{proof}

\section{``Yes''-instances of Ultimate Positivity}

\begin{prop}\label{Proposition: ``yes''-instances of UPP}
    A linear recurrence $(c,u) \in \R^{2n}$ is a robust ``yes''-instance of the Ultimate Positivity Problem if and only if it has a unique and simple dominant characteristic root which in addition is a positive real number whose coefficient in the exponential polynomial solution is a positive real number.
    Moreover, there exists an algorithm which takes as input a real linear recurrence and halts if and only if the linear recurrence is a robust ``yes''-instance of the Ultimate Positivity Problem.
\end{prop}
\begin{proof}
    Clearly any instance of the described form is a ``yes''-instance.
    We can apply the algorithm from Lemma \ref{Lemma: computing dominant characteristic real root and its coefficient}
    to check whether there exists a unique and simple dominant characteristic root which in addition is a positive real number,
    and if so compute its coefficient in the exponential polynomial solution.
    We can then semi-decide if this coefficient is positive.
    This shows that the set of instances of the described form is semi-decidable, which implies that any such instance is robust.

    It now remains to show that there are no further robust ``yes''-instances.
    Let $(c,u)$ be a ``yes''-instance of order $n$.
    Assume that $(c,u)$ is not of the described form.
    Then either $(c,u)$ does not have a unique and simple dominant characteristic root $\rho$ which in addition is a positive real number, or it does and the coefficient of $\rho^k$ in the exponential polynomial solution is non-positive.

    Assume first that $(c,u)$ has a unique and simple dominant characteristic root $\rho$ which in addition is a positive real number, but its coefficient $c$ in the exponential polynomial solution is non-positive.
    By Proposition \ref{Proposition: linear recurrence depends continuously on exponential polynomial solution} it suffices to show that there exist arbitrarily small perturbations of the exponential polynomial solution which fail to be ultimately positive.
    By an arbitrarily small perturbation of the formal exponential polynomial we can ensure that $\rho$ is the only dominant characteristic root 
    and that its coefficient is strictly negative.
    This slightly perturbed sequence will be negative for all large values of $k$.
    It follows that $(c,u)$ is not robust.

    It remains to consider the case where $(c,u)$ does not have a unique and simple dominant characteristic root $\rho$ which is a positive real number.
    We can further assume that the dominant characteristic roots of $(c,u)$ have strictly positive modulus. 
    Otherwise we have $c = 0$, which immediately implies that $(c,u)$ is not robust.
    Since we assume that $(c,u)$ is a ``yes''-instance it follows from Lemma \ref{Lemma: behaviour of non-positive dominant part} that $(c,u)$ has to have a dominant characteristic root $\rho$ which is a positive real number.
    Hence, either $\rho$ is not simple or there exists another dominant characteristic root $\lambda \in \C \setminus [0, +\infty)$.
    
    If there exists a dominant characteristic root $\lambda \in \C \setminus [0, +\infty)$ 
    then under arbitrarily small perturbations of the exponential polynomial solution the root $\lambda$ becomes strictly larger in modulus than $\rho$
    and its coefficient can be ensured to be non-zero.
    By Proposition \ref{Proposition: linear recurrence depends continuously on exponential polynomial solution} this induces an arbitrarily small perturbation of the instance $(c,u)$.
    It follows from Lemma \ref{Lemma: behaviour of non-positive dominant part} that the perturbed instance is a ``no''-instance of the Positivity Problem.
    Hence $(c,u)$ is not robust.

    Finally assume that $\rho$ is not simple.
    Let $D \subseteq \C^n$ and $h \colon D \to \R^n$ be as in Proposition \ref{Proposition: coefficients depend continuously on characteristic roots}.
    Choose $p \in D$ with $h(p) = c$.
    By continuity of $h$ a small perturbation of $p$ within $D$ induces a small perturbation of the coefficients of $c$ and therefore a small perturbation of 
    the input $(c,u)$.
    Since $\rho$ is not simple the vector $p$ contains at least two entries equal to $\rho$.
    By an arbitrarily small perturbation of $p$ we can make these two entries into two complex conjugate numbers whose modulus is strictly larger than $\rho$.
    By an arbitrarily small perturbation of the exponential polynomial solution of the resulting perturbed instance we can further ensure that the coefficients of these complex conjugate characteristic roots in the exponential polynomial solution are non-zero.
    It follows from Proposition \ref{Proposition: linear recurrence depends continuously on exponential polynomial solution} that this induces an arbitrarily small perturbation of the original instance. 
    By Lemma \ref{Lemma: behaviour of non-positive dominant part} the resulting perturbed instance is not ultimately positive.
\end{proof}

\section{Positivity}

\begin{lem}\label{Lemma: bounding the rest}
    Let $(c,u)$ be a linear recurrence.
    Assume that $(c,u)$ has a unique and simple dominant characteristic root $\rho$, which in addition is a positive real number.
    Let $a \in \R$ be the coefficient of $\rho^k$ in the exponential polynomial solution.
    Then we can compute an index $N \in \N$ such that 
    \[
        |a| \rho^k > |u_k - a \rho^k|
    \]
    for all $k \geq N$.
\end{lem}
\begin{proof}
    By Lemma \ref{Lemma: computing dominant characteristic real root and its coefficient} we can compute $\rho$ and its coefficient $a$ in the exponential polynomial solution uniformly in $(c,u)$ subject to the promise that $(c,u)$ is of the required form.

    We can compute a real number $M$ such that $|\lambda| < M < \rho$ for all characteristic roots $\lambda$ of $(c,u)$ with $\lambda \neq \rho$.
    We can then compute an integer $p$ such that 
    \begin{equation}\label{eq: Positivity Problem Lemma Eq 1}
        (n - 1) \left(\frac{M}{\rho}\right)^p \left(1 + \left(\frac{M}{\rho}\right)^p\right)^{n - 2}  < \tfrac{1}{2}.
    \end{equation}
    
    For all $q = 0,\dots, p - 1$
    we can thus compute a linear recurrence of order $n - 1$ which generates the sequence 
    \begin{equation}\label{eq: bounding rest eq 1}
        v^{(q)}_{k} = \frac{u_{pk + q}}{\rho^{pk + q}} - a.
    \end{equation}
    Its characteristic roots are of the form $(\mu / \rho)^p$, where $\mu$ is a characteristic root of $(c,u)$ which is distinct from $\rho$.
    Letting $\mu_1,\dots, \mu_{n - 1}$ denote the characteristic roots, counted with multiplicity, the coefficients of the linear recurrence are given by 
    $e_1(\mu_1,\dots,\mu_{n-1})$, $\dots$, $e_{n - 1}(\mu_1,\dots,\mu_{n - 1})$,
    where 
    \[ 
        e_j(x_1,\dots,n-1) = \sum_{1 \leq k_1 \leq \dots \leq k_j \leq n} x_{k_1} \cdot \dots \cdot x_{k_j}
    \]
    is the $j^{\text{th}}$ elementary symmetric polynomial in $n - 1$ variables.
    Using the estimate $|\mu_j| < \left(\tfrac{M}{\rho}\right)^p$ we obtain:
    \[
        |e_j(\mu_1,\dots,\mu_{n-1})| \leq \binom{n - 1}{j} \left(\frac{M}{\rho}\right)^{pj}.
    \]

    We can thus estimate 
    \begin{align*}
        \left|v^{(q)}_{k + 1}\right|
        &\leq 
        \left(|e_1(\mu_1,\dots,\mu_{n-1})| + \dots + |e_{n - 1}(\mu_1,\dots,\mu_{n-1})|\right) 
            \max
            \left\{
                \left|v^{(q)}_k\right|, \dots, \left|v^{(q)}_{k - n + 1}\right|
            \right\}\\
        &\leq
        \left(\sum_{j = 1}^{n - 1} \binom{n - 1}{j} \left(\frac{M}{\rho}\right)^{pj}\right)
            \max
            \left\{
                \left|v^{(q)}_k\right|, \dots, \left|v^{(q)}_{k - n + 1}\right|
            \right\}\\
        &\leq (n - 1) \left(\frac{M}{\rho}\right)^p \left(\sum_{j = 0}^{n - 2} \binom{n - 2}{j} \left(\frac{M}{\rho}\right)^{pj}\right)
            \max
            \left\{
                \left|v^{(q)}_k\right|, \dots, \left|v^{(q)}_{k - n + 1}\right|
            \right\}\\
            &\leq
            (n - 1) \left(\frac{M}{\rho}\right)^p \left(1 + \left(\frac{M}{\rho}\right)^p\right)^{n - 2} 
            \max
            \left\{
                \left|v^{(q)}_k\right|, \dots, \left|v^{(q)}_{k - n + 1}\right|
            \right\}\\
        &< \frac{1}{2} 
            \max
            \left\{
                \left|v^{(q)}_k\right|, \dots, \left|v^{(q)}_{k - n + 1}\right|
            \right\}.
    \end{align*}
    The last inequality follows from \eqref{eq: Positivity Problem Lemma Eq 1}.
    By induction it follows for all $k > 1$ that 
    \[
        \left|v^{(q)}_{n - 1 + k}\right|
        < \frac{1}{2^k}
        \max
        \left\{
            v^{(q)}_{n - 1}, \dots, v^{(q)}_{1}
        \right\}.
    \]
    We can hence compute an index $N$ such that for all $q = 0,\dots, p - 1$
    and all $j \geq \tfrac{N}{p} - 1$ we have 
    $\left|v^{(q)}_{j}\right| < |a|$.
    Using \eqref{eq: bounding rest eq 1} we obtain 
    $\left|\tfrac{u_k}{\rho^k} - a\right| < |a|$
    for all $k \geq N$.
    Multiplication of this inequality with $\rho^k$ yields the result.
\end{proof}

\begin{prop}\label{Proposition: ``yes''-instances of Positivity}
    A ``yes''-instance $(c,u)$ of the Positivity Problem is robust if and only if it is a robust ``yes''-instance of the Ultimate Positivity Problem and satisfies 
    $u_k > 0$ for all $k$.
    Moreover, there exists an algorithm which takes as input a linear recurrence and halts if and only if it is a robust ``yes''-instance of the Positivity Problem.
\end{prop}
\begin{proof}
    Since Positivity implies Ultimate Positivity it is clear that any robust ``yes''-instance of the Positivity Problem must be a robust ``yes''-instance of the Ultimate Positivity Problem. 
    We claim that if $(c,u)$ is a ``yes''-instance of the Positivity Problem with $u_k = 0$ for some $k$, then there exists an arbitrarily small perturbation of the instance with $u_k < 0$.
    Indeed, we can choose $k$ to be minimal with this property.
    If $u_k$ is an initial value then we can slightly perturb $u_k$ to make it negative.
    Otherwise we have  
    $u_k = c_1 u_{k - 1} + \dots + c_n u_{k - n}$
    with $u_{k - j} > 0$ for $j = 1,\dots,n$ by minimality of $k$.
    Then an arbitrarily small perturbation of $c_1$, say, will make $u_k$ negative.
    This proves the claim.

    Now consider the following algorithm: given an instance $(c,u)$ of the Positivity Problem, semi-decide if it has a positive real dominant characteristic root $\rho$ which is simple and strictly larger than the modulus of any other characteristic root.
    If this semi-decision procedure halts, compute the coefficient $a$ of $\rho^k$ in the exponential polynomial solution.
    This is possible thanks to Lemma \ref{Lemma: computing dominant characteristic real root and its coefficient}
    Test if $a > 0$.
    If this test halts then compute an index $N$ such that $a \rho^k > |u_k - a\rho^k|$ for all $k \geq N$.
    This is possible thanks to Lemma \ref{Lemma: bounding the rest}.
    Then test for all $k < N$ if $u_k > 0$.

    Together with the characterisation of robust ``yes''-instances of the Ultimate Positivity Problem given in Proposition \ref{Proposition: ``yes''-instances of Positivity} it is now clear that this algorithm halts if and only if $(c,u)$ is a robust ``yes''-instance of the Positivity Problem.
\end{proof}

\begin{thm}
    The Positivity Problem is maximally partially decidable.
\end{thm}
\begin{proof}
    It is obvious that all ``no''-instances of Positivity are robust and algorithmically recognisable.
    Proposition \ref{Proposition: ``yes''-instances of Positivity} establishes that all robust ``yes''-instances are algorithmically recognisable.
\end{proof}

\section{The Skolem Problem}

\begin{prop}
    No ``yes''-instance of the Skolem Problem is robust.
\end{prop}
\begin{proof}
    Let $(c,u)$ be a ``yes''-instance of the Skolem Problem.
    Let us assume that $(c,u)$ has a dominant characteristic root $\lambda$ with non-zero imaginary part.
    An analogous but simpler proof establishes the claim in case that there exists a real dominant characteristic root.

    Since the vector of coefficients $c$ depends continuously on the vector of characteristic roots
    we can assume, up to slightly perturbing $(c,u)$,
    that $\lambda$ is simple, that $\lambda$ and $\bar{\lambda}$ are the only dominant characteristic roots, and that $\lambda/|\lambda|$ is a root of unity.
    The exponential polynomial solution of $(c,u)$ is then of the form 
    $c \lambda^k + \bar{c} \bar{\lambda}^k + r(k)$,
    with $|r(k)| < a b^{k}$ for some $a \geq 0$ and $b < M$.
    Writing $\lambda = M e^{i\varphi}$ and $c = \alpha + i \beta$ we have 
    \[
    c \lambda^k + \bar{c} \bar{\lambda}^k = 2M\left(\alpha \cos(k\varphi) - \beta \sin(k\varphi) \right).
    \]
    Thus, $c \lambda^k + \bar{c} \bar{\lambda}^k$ does not vanish so long as the point $(\alpha,\beta)$ lies outside 
    the set 
    $S = \Set{(x,y) \in \R^2}{\exists k. x \cos(k\varphi) - y \sin(k\varphi) = 0}$.
    Since $e^{i\varphi}$ is a root of unity the expressions $\cos(k\varphi)$ and $\sin(k\varphi)$ admit only finitely many different values for $k \in \N$. 
    Additionally, the expressions $\cos(k\varphi)$ and $\sin(k\varphi)$ never vanish simultaneously.
    Therefore the set $S$ is a finite union of straight lines in $\R^2$.
    It follows that there exist arbitrarily small perturbations of $\alpha$ and $\beta$ such that $\alpha \cos(k\varphi) - \beta \sin(k\varphi)$ never vanishes.
    Since this expression only admits finitely many values it follows that 
    $|u_k| > \varepsilon M^k - a b^k$
    with $\varepsilon > 0$ and $b < M$.

    In particular up to arbitrarily small perturbation $(c,u)$ has only finitely many zeroes $k_1 < k_2 < \dots < k_m$.
    As above we can find arbitrarily small $\gamma$ and $\delta$ such that 
    $(\gamma + i\delta) \lambda^{k_j} + (\gamma - i\delta)\bar{\lambda}^{k_j}$
    is non-zero for $j = 1,\dots,m$.
    Replace the coefficient $c$ in the exponential polynomial solution of $(c,u)$ by 
    $c + (\gamma + i\delta)$ and $\bar{c}$ with $\bar{c} + (\gamma - i\delta)$.
    Then the zeroes $k_1,\dots,k_m$ are removed by construction.
    By choosing $\gamma$ and $\delta$ sufficiently small we can ensure that no new zeroes are added.
    By Proposition \ref{Proposition: linear recurrence depends continuously on exponential polynomial solution} this induces an arbitrarily small perturbation of the instance $(c,u)$.
\end{proof}

\begin{prop}
    A ``no''-instance of the Skolem Problem is robust if and only if one of the two following conditions is met:
    \begin{enumerate}
        \item 
        It has a simple real dominant characteristic root $\rho$ 
        with $|\rho| > |\mu|$ for all characteristic roots $\mu \neq \rho$,
        and the coefficient of $\rho$ in the exponential polynomial solution is non-zero.
        \item 
        It has exactly two dominant characteristic roots, both of which are simple and real,
        and their respective coefficients in the exponential polynomial solution are non-zero
        and have different absolute values.
    \end{enumerate}
    Moreover, there exists an algorithm which takes as input a linear recurrence $(c,u)$ and halts if and only if $(c,u)$ is a robust ``no''-instance of the Skolem Problem.
\end{prop}
\begin{proof}
    Assume that a ``no''-instance $(c,u)$ of the Skolem Problem has a complex dominant characteristic root $\lambda \in \C \setminus \R$.
    By Proposition \ref{Proposition: coefficients depend continuously on characteristic roots} a small perturbation of the characteristic roots induces a small perturbation of the instance $(c,u)$,
    so that we can assume up to an arbitrarily small perturbation of $(c,u)$ 
    that $\lambda$ is simple, not a root of unity, and that that $\lambda$ and $\bar{\lambda}$ are the only dominant characteristic roots of $(c,u)$.
    Let $M = |\lambda|$.
    It follows from Dirichlet's approximation theorem that the sequence $(\lambda/M)^k$ is dense in the unit circle $S^1 \subseteq \C$.
    In particular, for every $\varepsilon > 0$ there exist infinitely many $k \in \N$ such that 
    $|(\lambda/M)^k| < \varepsilon$.
    If follows that there exists an index $K$ such that  
    $|u_K/M^K| < \varepsilon$.
    Now, the linear recurrence sequence 
    \[
        w_k = u_k - \frac{u_K}{2\lambda^K} \lambda^k - \frac{u_K}{2\bar{\lambda}^K} \bar{\lambda}^k
    \] 
    satisfies $w_K = 0$.
    It is represented by a formal exponential polynomial which is $\varepsilon$-close to the exponential polynomial solution of $(c,u)$.
    Since $\varepsilon$ can be chosen to be arbitrarily small, this yields an arbitrarily small perturbation of $(c,u)$ which is a ``yes''-instance of the Skolem Problem.
    It follows that $(c,u)$ is not robust.

    If $(c,u)$ has a real dominant characteristic root which is not simple then a small perturbation of $(c,u)$ has a complex dominant characteristic root.
    It equally follows that $(c,u)$ is not robust.

    Thus, if $(c,u)$ is a robust ``no''-instance of the Skolem Problem then its dominant characteristic roots are all simple and real.
    Assume that $(c,u)$ has only one dominant characteristic root $\rho$.
    Let $M$ be its absolute value.
    Note that $M > 0$ since we assume that the instance is a ``no''-instance.
    If the coefficient of $\rho^k$ in the exponential polynomial solution is zero, then for all $\varepsilon > 0$ there exists an index $K$ such that 
    $|u_K/M^K| < \varepsilon$.
    Then the linear recurrence sequence $w_k = u_k - \frac{u_K}{\rho^K} \rho^k$ has a zero and is represented by a formal exponential polynomial which is $\varepsilon$-close to the exponential polynomial solution of $(c,u)$.
    It follows that $(c,u)$ is not robust.
    On the other hand, if the coefficient of $\rho^k$ is non-zero then it is easy to see that the instance is a robust ``no''-instance.
    
    Now assume that $(c,u)$ has exactly two dominant characteristic roots $\rho > 0$ and $-\rho < 0$.
    If the coefficient in the exponential polynomial solution of $\rho^k$, say, is zero then by an arbitrarily small perturbation we can ensure that $\rho$ is the only dominant characteristic root of $(c,u)$ and it follows that $(c,u)$ is not robust.
    Hence the coefficients of $\rho^k$ and of $(-\rho)^k$ must be non-zero.
    If they have the same absolute value then the formal dominant part of the exponential polynomial solution vanishes either for all odd indexes or for all even indexes.
    Again it follows that for all $\varepsilon > 0$ there exists $K$ such that $|u_K/\rho^K| < \varepsilon$ and by the same argument as before it follows that $(c,u)$ is not robust.
    On the other hand, if the coefficients of $\rho^k$ and $(-\rho)^k$ are non-zero and have different absolute values then it is easy to see that the resulting instance is a robust ``no''-instance of the Skolem Problem.
    This concludes the characterisation of the robust ``no''-instances.

    To semi-decide if a given instance $(c,u)$ is a ``no''-instance, run the following two tests in parallel:
    \begin{enumerate}
        \item $(c,u)$ has a simple real dominant characteristic root with $|\rho| > |\mu|$ for all characteristic roots $\mu \neq \rho$,
        and the coefficient of $\rho$ in the exponential polynomial solution is non-zero.
        \item $(c,u)$ has a simple positive real characteristic root $\rho_+$ 
        and a simple negative real characteristic root $\rho_{-}$
        such that $|\rho_+| > |\mu|$ and $|\rho_{-}| > |\mu|$ for all characteristic roots $\mu \notin \{\rho_{+},\rho_{-}\}$.
        The coefficients of $\rho_{+}$ and $\rho_{-}$ in the exponential polynomial solution are distinct and both non-zero.
    \end{enumerate}
    That the first test is effective follows essentially from Lemma \ref{Lemma: computing dominant characteristic real root and its coefficient}.
    The second test can be carried out effectively by similar ideas:
    Choose a rational number $\varepsilon > 0$.
    Compute a list of rational boxes $B_1,\dots,B_s \subseteq \C$ of width $\varepsilon$ such that each box is guaranteed to contain a complex root of the characteristic polynomial and each root of the characteristic polynomial is contained in a box.
    Test if there exist boxes $B_+$ and $B_{-}$ which intersect the real axis and contain a unique root,
    such that all real numbers contained in $B_{+}$ are positive,
    all real numbers contained in $B_{-}$ are negative,
    the modulus of all numbers contained in $B_{+}$ is strictly larger than the modulus of all numbers contained in boxes other than $B_{+}$ and $B_{-}$,
    and the same is true for the modulus of all numbers contained in $B_{-}$.
    Clearly this can be tested effectively in finite time.
    If this test does not succeed then retry with $\varepsilon/2$ replacing $\varepsilon$.
    If the test succeeds then there is a unique characteristic root $\rho_+ \in B_+$ and a unique characteristic root $\rho_- \in B_-$, both of which are simple and real.
    We can effectively compute the respective coefficients of $\rho_+$ and $\rho_-$ in the exponential polynomial solution of $(c,u)$ 
    thanks to Corollary \ref{Corollary: computing coefficient of eigenvalue with known multiplicity}
    and test if they are distinct and non-zero. 

    Whenever one of the tests terminates we can effectively compute positive real numbers $\delta$ and $r$ such that the absolute value of the formal dominant part of the exponential polynomial solution is bounded from below by $\delta r^k$.
    It then follows as in the proof of Lemma \ref{Lemma: bounding the rest} that we can compute an index $K \in \N$ such that $|u_k| > 0$ for all $k \geq K$.
    To verify that the given instance is a ``no''-instance it hence suffices to verify that $u_k > 0$ for all $k < K$.

    It is clear that if this algorithm halts then the given instance is a ``no''-instance.
    Conversely, if the instance is a robust ``no''-instance then it meets one of the two criteria above.
    If it meets the fist criterion then the first test will terminate.
    If it meets the second criterion then the second test will terminate.
\end{proof}

\section{Approximate root clusterings and possible root configurations}

Finally we turn to the problem of computably recognising robust ``no''-instances of Ultimate Positivity.
This will be considerably more involved than the previous results, and further preparatory work is required.
The ideas we introduce here have been motivated in Example \ref{Example: proof idea ultimate positivity} in the introduction.

Let $P \in \R[x]$ be a non-constant univariate real polynomial.
An \emph{approximate root clustering} for $p$ is a finite list 
\[
    \left\langle (B_1, N_1), \dots, (B_s, N_s) \right\rangle
\]
where each $B_j \subseteq \C$ is a rational box, \textit{i.e.}, a product of intervals with rational endpoints, and each $N_j$ is a positive integer, such that 
\begin{enumerate}
    \item Every complex root of $P$ is contained in one of the boxes $B_j$.
    \item For all $j = 1,\dots,s$ the number $N_j$ is the number of roots of $p$ in $B_j$ counted with multiplicity.
    \item If a $B_j$ intersects the real line then its reflection about the real line is disjoint from all boxes $B_k$ with $k \neq j$.
    \item If a $B_j$ does not intersect the real line then there exists a unique $B_k$ with $k \neq j$ and $\bar{B}_j \cap B_k \neq \emptyset$,
    where $\bar{B}_j$ denotes the reflection of $B_j$ about the real axis.
    \item There exists an index $0 \leq a \leq s$ such that the boxes $B_1,\dots,B_a$ intersect the real line and no box $B_j$ with $j > a$ intersects the real line.
    \item For all $1 \leq j < a$ with $a$ as in the previous item, we have 
    $\min \left(B_j \cap \R\right) > \max\left(B_{j + 1} \cap \R\right)$.
    \item If $a \neq s$ then there exists and index $a < b \leq s$ such that the boxes $B_{a + 1},\dots,B_b$ are contained in the upper half-plane and the boxes $B_j$ with $j > b$ are contained in the lower half-plane. 
\end{enumerate}

The elements $(B_j,N_j)$ of an approximate root clustering are called \emph{clusters}.
A $(B_j, N_j)$ where $B_j$ intersects the real line is called a \emph{real cluster}.
Otherwise it is called a \emph{complex cluster}.
If $(B_j, N_j)$ is a complex cluster and $(B_k, N_k)$ is the unique cluster with $k \neq j$ and $\bar{B}_j \cap B_k \neq \emptyset$
then we call $(B_j,N_j)$ and $(B_k,N_k)$ \emph{complex conjugate clusters} and write 
$(B_j,N_j) = \overline{(B_k,N_k)}$.
Note that it follows from the definition that for complex conjugate clusters $(B_j,N_j)$ and $(B_k,N_k)$ we have $N_j = N_k$.

Let $(B_j,N_j)$ be a real cluster.
A \emph{possible root configuration} for $(B_j,N_j)$ is a pair of lists 
\[
    \langle (\rho_{1}, r_1),\dots, (\rho_{d}, r_{d}) \rangle,
    \langle (\lambda_{1}, m_1),\dots, (\lambda_e, m_e) \rangle
\]
where $\rho_1,\dots,\rho_d$ and $\lambda_1,\dots,\lambda_e$ are distinct variables
and $r_1,\dots,r_d, m_1,\dots,m_e$ are positive integers with 
$r_1 + \dots + r_d + 2m_1 + \dots + 2m_e = N_j$.
For every $j = 1,\dots, d$ the number $r_j$ is called the \emph{multiplicity} of the variable $\rho_j$
and for $j = 1,\dots, e$ the number $m_j$ is called the multiplicity of the variable $\lambda_j$.
The intention is that the variables $\rho_1,\dots,\rho_d$ represent real roots with multiplicities $r_1,\dots, r_d$
and $\lambda_1,\dots,\lambda_e$ represent complex roots with positive imaginary part and multiplicities $m_1,\dots,m_e$.

Let $(B_j,N_j)$ be a complex cluster.
A \emph{possible root configuration} for $(B_j,N_j)$ is a list 
\[
    \langle (\lambda_{1}, m_1),\dots, (\lambda_e, m_e) \rangle
\]
where $\lambda_1,\dots,\lambda_e$ are distinct variables
and $m_1,\dots,m_e$ are positive integers
with $m_1 + \dots + m_e = N_j$.
Again, for $j = 1,\dots, e$ the number $m_j$ is called the multiplicity of the variable $\lambda_j$.

Now let 
$\left\langle (B_1, N_1), \dots, (B_s, N_s) \right\rangle$
be an approximate root clustering.
Let $a$ be the last index of a real cluster and $b$ be the last index of a cluster in the upper half-plane.
A \emph{possible root configuration} for this root clustering is a list 
$
    \langle
        R_1,\dots, R_a, C_1, \dots, C_{b - a}
    \rangle 
$
where $R_1,\dots, R_a$ are possible root configurations for the real clusters
$(B_1, N_1), \dots, (B_a, N_a)$
and 
$C_1,\dots, C_{b - a}$
are possible root configurations for the complex clusters
$(B_{a + 1}, N_{a + 1}), \dots, (B_{b}, N_b)$,
such that all variables occurring in the possible root configurations for different clusters are distinct.
We call the variables of the form $\rho_j$ the \emph{real variables} and the variables of the form $\lambda_j$ the \emph{complex variables} of the possible root configuration.
Formally, a real variable is a variable that occurs in the first list associated with some $R_j$ and any other variable is a complex variable.

\begin{prop}\label{Proposition: computing approximate root clustering}
    Given a non-constant real polynomial $P \in \R[x]$, encoded as a vector $(a_n,\dots,a_0) \in \R^{n + 1}$ where 
    $p(x) = a_n x^n + \dots + a_0$
    and $a_n \neq 0$,
    and a natural number $p \in \N$ 
    we can compute an approximate root clustering 
    \[\left\langle (B_1, N_1), \dots, (B_s, N_s) \right\rangle\]
    for $P$ such that each $B_j$ has width at most $2^{-p}$.
\end{prop}
\begin{proof}[Proof Sketch]
    By Theorem \ref{Theorem: Specker Zeroes} we can compute a complex vector $(\mu_1,\dots,\mu_n) \in \C^{n}$ which contains all roots of $P$
    such that a root of multiplicity $m$ occurs precisely $m$ times in this vector.
    Choose a small rational number $\delta > 0$.
    Approximate each $\mu_j$ to accuracy $\delta/2$ and put a rational box $C_j$ of width $\delta$ around this rational approximation.
    Arrange these boxes into clusters of the form $C_{j_1},\dots,C_{j_k}$ such that 
    the unions $C_{j_1} \cup \dots \cup C_{j_k}$ are connected and each $C_{j_i}$
    is disjoint from all boxes $C_k$ with $k \notin \{j_1,\dots,j_k\}$.
    Initialise an empty list $L = \langle\rangle$.
    For each such cluster compute a rational box $B$ that contains 
    $C_{j_1} \cup \dots \cup C_{j_k}$
    and add the element $(B, k)$ to the list $L$.
    The boxes $B$ in the list $L$ may have width larger than $2^{-p}$
    and the resulting list $L$ may not be an approximate root clustering since the reflection of a real cluster about the real line may intersect another cluster,
    or the reflection of a complex cluster about the real line may intersect more than one other cluster.
    If either of these cases occurs repeat the algorithm with $\delta/2$ replacing $\delta$.
    It is easy to see that for sufficiently small $\delta$ this algorithm will produce an approximate root clustering as desired.
\end{proof}

\begin{prop}\label{Proposition: computing all possible root configurations}
    Given an approximate root clustering $\left\langle (B_1, N_1), \dots, (B_s, N_s) \right\rangle$
    we can compute a list containing all possible root configurations for this clustering.
\end{prop}

Let $P \in \R[x]$ be a non-constant real polynomial.
Let $\mathcal{C} = \left\langle (B_1, N_1), \dots, (B_s, N_s) \right\rangle$ be an approximate root clustering.
Let $\mathcal{R}$ be a possible root configuration for $\mathcal{C}$.
Let $\rho_1,\dots,\rho_d$ be its real variables and let $\lambda_1,\dots,\lambda_e$ be its complex variables.
Let $r_1,\dots,r_d$ and $m_1,\dots,m_e$ be their respective multiplicities.
Introduce new variables $\bar{\lambda}_1,\dots,\bar{\lambda}_e$.
The \emph{characteristic polynomial} of $\mathcal{R}$ is the polynomial 
\[
    (x - \rho_1)^{r_1}\cdot \dots \cdot (x - \rho_d)^{r_d} 
    \cdot 
    (x - \lambda_1)^{m_1} \cdot (x - \bar{\lambda}_1)^{m_1}
    \cdot \dots \cdot 
    (x - \lambda_e)^{m_e} \cdot (x - \bar{\lambda}_e)^{m_e}
\]
with coefficients in 
$\Z[x, \rho_1,\dots,\rho_d, \lambda_1,\bar{\lambda}_1,\dots,\lambda_e, \bar{\lambda}_e]$.
Writing this polynomial in the form 
\[
    x^n - c_1 x^{n - 1} - \dots - c_{n - 1} x - c_n
\]
we obtain a linear recurrence relation 
\[
    u_{k + 1} = c_1 u_k + \dots + c_n u_{k + 1 - n}
\]
with coefficients 
$c_j \in \Z[\rho_1,\dots,\rho_d, \lambda_1,\bar{\lambda}_1,\dots,\lambda_e, \bar{\lambda}_e]$.
This yields a ``formal'' linear recurrence $(c,u)$, where $u_1,\dots,u_n$ are fresh variables.
Call this the \emph{linear recurrence associated with $\mathcal{R}$}.
Call its companion matrix the companion matrix of $\mathcal{R}$ and call its exponential polynomial solution the exponential polynomial solution of $\mathcal{R}$.
The exponential polynomial solution of $\mathcal{R}$ is a term of the form 
\begin{align*}
    &\sum_{j = 1}^d \phi_j(k, u,\rho_1,\dots,\rho_d, \lambda_1,\bar{\lambda}_1,\dots,\lambda_e, \bar{\lambda}_e) \rho_j^k\\
    +
    &\sum_{j = 1}^e \psi_j(k, u,\rho_1,\dots,\rho_d, \lambda_1,\bar{\lambda}_1,\dots,\lambda_e, \bar{\lambda}_e) \lambda_j^k\\
    +
    &\sum_{j = 1}^e \bar{\psi}_j(k, u,\rho_1,\dots,\rho_d, \lambda_1,\bar{\lambda}_1,\dots,\lambda_e, \bar{\lambda}_e) \bar{\lambda}_j^k,
\end{align*}
where the $\phi_j$'s and $\psi_j$'s are rational functions with rational number coefficients.
It is clearly computable from $\mathcal{R}$.

Let $\mathcal{R}$ be a possible root configuration for an approximate root clustering $\mathcal{C}$.
Let $(B_{a + 1},N_{a + 1}), \dots, (B_{b},N_b)$
be the list of complex clusters in $\mathcal{C}$ in the upper half plane.
Let $(B_1,N_1),\dots,(B_a,N_a)$ be the list of real clusters in $\mathcal{C}$.
Then $\mathcal{R}$ consists of possible root configurations $R_1,\dots, R_a, C_1, \dots, C_{b - a}$ for the individual clusters.
Let $\rho_1,\dots, \rho_d$ be the real variables which occur in $\mathcal{R}$ 
and let $\lambda_1,\dots,\lambda_e$ be the complex variables.
Assume without loss of generality that there exists an integer $w$ such that  
$\lambda_1,\dots,\lambda_w$ belong to root configurations for real clusters 
and that $\lambda_{w + 1},\dots,\lambda_e$ belong to root configurations for complex clusters.
Let 
$\beta \colon \{1,\dots,d\} \to \{1,\dots,a\}$
be the function that assigns to an integer $j$ the root configuration $R_{\beta(j)}$ in which the variable $\rho_j$ occurs.
Let 
$\gamma \colon \{1,\dots,w\} \to \{1,\dots,a\}$
be the function that assigns to an integer $j$ the root configuration $R_{\gamma(j)}$ in which the variable $\lambda_j$ occurs.
Let 
$\delta \colon \{w + 1,\dots, e\} \to \{1,\dots,b - a\}$
be the function that assigns to an integer $j$ the root configuration $C_{\delta(j)}$ in which the variable $\lambda_j$ occurs.
The \emph{domain} $D_{\mathcal{R}} \subseteq \R^d \times \C^e$ of $\mathcal{R}$ is the set of all vectors 
$(x_1,\dots,x_d, z_1,\dots,z_e) \in \R^d\times \C^e$ 
such that 
$x_1 > \dots > x_d$,
$z_j \neq z_k$ for $j \neq k$,
$\Im(z_j) > 0$ for all $j = 1, \dots, e$,
$x_j \in B_{\beta(j)}$ for all $j = 1,\dots, d$,
$z_j \in B_{\gamma(j)}$ for all $j = 1,\dots,w$,
and 
$z_j \in B_{a + \delta(j)}$ for all $j = w + 1,\dots, e$.

Thus, the domain of $\mathcal{R}$ is the set of all ``valid assignments'' to the variables $\rho_1,\dots,\rho_d$, and $\lambda_1,\dots,\lambda_e$.
To any point 
$(\alpha, u) \in D_{\mathcal{R}} \times \R^n$ 
corresponds a real linear recurrence which is obtained by substituting $\alpha$ for 
$\rho_1,\dots,\rho_d, \lambda_1,\dots,\lambda_e$
and $u$ for the variables representing initial values in the linear recurrence associated with $\mathcal{R}$.
We call this the \emph{linear recurrence associated with $\alpha$ with initial values $u$}.

We conclude with two obvious observations:

\begin{lem}\label{Lemma: points in the domain of a root clustering are small perturbations of the linear recurrence}
    There exists a computable function 
    $\varphi \colon \coprod_{n \in \N} \R^{2n} \times (0,1) \to (0,+\infty)$
    with the following property:

    Let $(c,u)$ be a linear recurrence with characteristic polynomial $P$.
    Let $\mathcal{C}$ be an approximate root clustering for $P$ to accuracy $\varepsilon > 0$.
    Let $\mathcal{R}$ be a possible root configuration for $\mathcal{C}$.
    Let $\alpha \in D_{\mathcal{R}}$ and let $v \in \R^n$ with $|v - u| < \varepsilon$.
    Then the linear recurrence associated with $\alpha$ with initial values $v$
    is 
    $\varphi(c,u,\varepsilon)$-close to $(c,u)$.
    Moreover, for all $(c,u) \in \coprod_{n \in \N} \R^{2n}$
    we have  
    $\varphi(c,u,\varepsilon) \to 0$ as $\varepsilon \to 0$.
\end{lem}

\begin{lem}
    Let $\mathcal{C}$ be an approximate root clustering for a polynomial $P$.
    Let $\mathcal{R}$ be a possible root configuration for $\mathcal{C}$.
    Then the domain $D_{\mathcal{R}}$ is definable in the first-order theory of the reals 
    as a subset of $\R^e\times \R^{2d}$.
\end{lem}

\section{``No''-instances of Ultimate Positivity}

\begin{thm}\label{Theorem: robust ``no''-instances of Ultimate Positivity}
    There exists an algorithm which takes as input a linear recurrence $(c,u)$ and halts if and only if $(c,u)$ is a robust ``no''-instance of the Ultimate Positivity Problem.
\end{thm}
\begin{proof}
Consider the following algorithm.

For all $p \in \N$ do the following:
\begin{enumerate}
    \item Compute an approximate root clustering $\mathcal{C}$ of the characteristic polynomial of $(c,u)$ to accuracy $2^{-p}$.
    \item Compute a rational box $B$ of width $2^{-p}$ that contains the vector $u$ of initial values.
    \item For all possible root configurations $\mathcal{R}$ of $\mathcal{C}$ do the following:
    \begin{enumerate}
        \item Symbolically compute the exponential polynomial solution of the linear recurrence associated with $\mathcal{R}$.
        \item Use the symbolic exponential polynomial solution to construct a sentence in the first-order theory of the reals 
        which expresses that for all points 
        $\alpha \in D_{\mathcal{R}}$
        and all 
        $v \in B$
        the formal leading coefficient of the largest real characteristic root in the exponential polynomial solution of the linear recurrence associated with 
        $\alpha$ with initial values $v$ is negative 
        or there exists a characteristic root $\lambda \in \C \setminus [0,+\infty)$ 
        whose modulus is larger than any positive real characteristic root such that the coefficient of $\lambda$ is non-zero.
        If there exist no real roots then only include the second condition.
        \item Employ the Tarski-Seidenberg theorem to decide whether the above sentence is true.
        If it is true continue with the next root configuration, or leave the loop if there are no root configurations left.
        If the sentence is false break out of the loop and continue with the next $p$.
    \end{enumerate}
    \item If all sentences constructed in the above loop are true, then halt.
\end{enumerate}

We claim that this algorithm halts on a given instance $(c,u)$ if and only if the instance is a robust ``no''-instance.
By construction if the algorithm halts then the instance $(c,u)$ either has no positive real characteristic roots or the formal leading coefficient of the largest positive real root $\rho$ in the exponential polynomial solution of $(c,u)$ is negative, or there exists a characteristic root $\lambda \in \C \setminus [0,+\infty)$ with $|\lambda| > \rho$
such that the coefficient of $\lambda$ is non-zero.
It follows from Lemma \ref{Lemma: behaviour of non-positive dominant part} that $(c,u)$ is a ``no''-instance of Ultimate Positivity.

Conversely, assume that $(c,u)$ is a robust ``no''-instance.
Suppose for the sake of contradiction that the algorithm does not halt.
Then by Lemma \ref{Lemma: points in the domain of a root clustering are small perturbations of the linear recurrence} there exists an arbitrarily small perturbation $(d,v)$ of $(c,u)$ such that either $(d,v)$ has no positive real characteristic roots and the coefficients of all characteristic roots are zero or the formal leading coefficient of the largest positive real characteristic root $\rho$ in the exponential polynomial solution of $(d,v)$ is non-negative and the coefficient of every complex or negative characteristic root with strictly larger modulus is zero.
If the former is the case then the sequence generated by $(d,v)$ is identically equal to zero and hence a ``yes''-instance of Ultimate Positivity.
Let us now assume that $(d,v)$ has a positive real characteristic root.
By Proposition \ref{Proposition: linear recurrence depends continuously on exponential polynomial solution} a small perturbation of the exponential polynomial solution of $(d,v)$ induces a small perturbation of the linear recurrence $(d,v)$.
By an arbitrarily small perturbation of the exponential polynomial we can ensure that the largest positive characteristic root $\rho$ has a positive leading coefficient and that every complex or negative characteristic root with a non-zero coefficient has strictly smaller modulus than $\rho$.
The resulting instance is then clearly a ``yes''-instance of the Ultimate Positivity Problem.
It follows that the original instance is not robust.
\end{proof}

Unlike the previous results Theorem \ref{Theorem: robust ``no''-instances of Ultimate Positivity} falls short of providing a satisfactory characterisation of the robust ``no''-instances of Ultimate Positivity.
It is therefore worth pointing out that at least the robust simple instances of the Ultimate Positivity Problem admit a nice characterisation.
A simple instance is one whose characteristic roots are all simple.
With the help of Corollary \ref{Corollary: computing coefficient of eigenvalue with known multiplicity} these instances are much easier to computably recognise than the general ones.

\begin{prop}\label{Proposition: simple robust instances of Ultimate Positivity}
    A simple instance of the Ultimate Positivity Problem is robust if and only if it satisfies one of the two following conditions:
    \begin{enumerate}
        \item It has a positive real dominant characteristic root $\rho$ with $|\rho| > |\lambda|$ for all other characteristic roots,
        and the coefficient of $\rho$ in the exponential polynomial solution is non-zero.
        If the coefficient is negative then it is a robust ``no''-instance.
        Otherwise it is a robust ``yes''-instance.
        \item It has a characteristic root $\lambda \in \C \setminus [0,\infty)$ with $|\lambda| > \rho$ for all positive characteristic roots $\rho$
        whose coefficient in the exponential polynomial polynomial is non-zero.
        In this case it is a robust ``no''-instance.
    \end{enumerate}
\end{prop}

\section{On the measure of the robust instances}

We observe that the simple ``yes''-instances of the Positivity Problem and the Ultimate Positivity Problem have full measure.
This is certainly false for the Skolem Problem, since the robust instances of the Skolem Problem are not even dense in $\R^{2n}$. 

\begin{prop}
    For all $n \in \N$ the sets of simple robust instances of the Positivity Problem and the Ultimate Positivity Problem each have full measure. 
\end{prop}
\begin{proof}
    Consider all spaces $R_{s,t} = \R^s \times \C^t$ with $s + 2t = n$.
    These can all be identified with the space $\R^n$.
    On each of these spaces we have a map 
    $h_{s,t} \colon R_{s,t} \to \R^n$ 
    which sends the vector $(\rho_1, \dots, \rho_s, \lambda_1, \dots, \lambda_t)$ to the $n$ lowest coefficients of the monic polynomial 
    \[
        (x - \rho_1) \cdot \dots \cdot (x - \rho_s) (x - \lambda_1) (x - \overline{\lambda_1}) \cdot \dots \cdot  (x - \lambda_t) (x - \overline{\lambda_t}).
    \]
    This map is clearly differentiable as a map $\R^n \to \R^n$.
    It hence maps null sets to null sets.
    Now, the sets
    \[
        A_{s,t} = \left\{ (\rho_1, \dots, \rho_s, \lambda_1, \dots, \lambda_t) \in \R^s \times \C^t \mid |\rho_i| = |\rho_j| \text{ for some } i \neq j \right\},
    \]
    \[
        B_{s,t} = \left\{ (\rho_1, \dots, \rho_s, \lambda_1, \dots, \lambda_t) \in \R^s \times \C^t \mid |\rho_i| = |\lambda_j| \text{ for some } i \neq j \right\},
    \]
    and
    \[
        C_{s,t} = \left\{ (\rho_1, \dots, \rho_s, \lambda_1, \dots, \lambda_t) \in \R^s \times \C^t \mid |\lambda_i| = |\lambda_j| \text{ for some } i \neq j \right\}
    \]
    all have measure zero.
    As there are only finitely many spaces $R_{s,t}$, we may remove the images $h_{s,t}(A_{s,t} \cup B_{s,t} \cup C_{s,t})$ from $\R^n$ and are still left with a set of full measure.
    This implies that the set of instances $(c, u) \in \R^{2n}$ whose characteristic roots are simple and distinct in modulus have full measure.

    Now consider the spaces $E_{s,t} \subseteq \R^{2s} \times \C^{2t}$ with $s + 2t = n$, 
    where a point 
    \[
    (\rho_1, \dots, \rho_s, \alpha_1, \dots, \alpha_s, \lambda_1, \dots, \lambda_t, \beta_1, \dots, \beta_t)
    \]
    is an element of 
    $E_{s,t}$ if and only if 
    $(\rho_1, \dots, \rho_s, \lambda_1, \dots, \lambda_t) \in R_{s,t} \setminus (A_{s,t} \cup B_{s,t} \cup C_{s,t})$.
    On each of these spaces we have a map 
    $m_{s,t} \colon E_{s,t} \to \R^{2n}$
    which sends the data 
    \[
        (\rho_1, \dots, \rho_s, \alpha_1, \dots, \alpha_s, \lambda_1, \dots, \lambda_t, \beta_1, \dots, \beta_t)
    \]
    to a code $(c_1, \dots, c_n, u_1, \dots, u_n)$ for the linear recurrence which defines the exponential polynomial
    \[
        \alpha_1 \rho_1^k + \dots + \alpha_s\rho_s^k + \beta_1 \lambda_1^k + \overline{\beta_1} \overline{\lambda_1}^k + \dots + \beta_t \lambda_t^k + \overline{\beta_t} \overline{\lambda_t}^k.
    \]
    As this map is again differentiable, it sends null sets to null sets. 
    The set 
    \begin{align*}
        D_{s,t} = 
        \big\{ 
            (\rho_1, \dots, \rho_s, \alpha_1, \dots, \alpha_s, &\lambda_1, \dots, \lambda_t, \beta_1, \dots, \beta_t) \in E_{s,t}
            \mid\\
            &\alpha_i = 0 \text{ for some } i 
            \text{ or }
            \beta_i = 0 \text{ for some } i
        \big\}
    \end{align*}
    is clearly a null set, so that its image in $\R^{2n}$ under $m_{s,t}$ is again a null set.
    It follows that we may remove all sets of the form $m_{s,t}(D_{s,t})$ from $\R^{2n}$ to be left with a set of full measure.

    In summary, the set of linear recurrences $(c, u)$ whose characteristic roots are all distinct and have non-zero coefficients in the exponential polynomial solution has full measure.
    All of these linear recurrences are robust instances of the Ultimate Positivity Problem.
    The remaining non-robust instances of the Positivity Problem are ``yes''-instances which have a zero.
    Within the space $E_{s,t}$ the linear recurrences which have a zero can be identified with the union of the sets 
    \begin{align*}
        Z_k = 
        \big\{ 
            &(\rho_1, \dots, \rho_s, \alpha_1, \dots, \alpha_s, \lambda_1, \dots, \lambda_t, \beta_1, \dots, \beta_t) \in E_{s,t}
            \mid\\
            &\alpha_1 \rho_1^k + \dots + \alpha_s\rho_s^k + \beta_1 \lambda_1^k + \overline{\beta_1} \overline{\lambda_1}^k + \dots + \beta_t \lambda_t^k + \overline{\beta_t} \overline{\lambda_t}^k = 0
        \big\}.
    \end{align*}
    Under the identification of $E_{s,t}$ with an open subspace of $\R^{2n}$, each $Z_k$ is the zero set of a differentiable real-valued function $f$.
    Its gradient does not vanish in $E_{s,t} \setminus D_{s,t}$, so that $Z_k \cap (E_{s,t} \cap D_{s,t})$ has measure zero.
    It follows that the remaining non-robust instances also have measure zero.
    Thus everything is shown. 
\end{proof}

\bibliographystyle{alpha}
\bibliography{dplrsr}
\nocite{*}

\end{document}